\documentclass[aps,preprint]{revtex4-1}

\usepackage{feynmp}
\usepackage{amsmath}
\usepackage{graphicx}
\usepackage{rotating}
\usepackage{multirow}
\usepackage{latexsym}
\usepackage{textcomp}
\usepackage{verbatim}
\usepackage{color}
\usepackage{pict2e}
\usepackage{bm}
\usepackage{subfigure}
\usepackage{keyval}
\usepackage{ragged2e}
\usepackage{dsfont}
\usepackage{amssymb}
\usepackage{enumitem}
\usepackage{pstricks}
\usepackage{epstopdf} 
\usepackage{mathtools}
\usepackage{hyperref}

\newcommand{\osum}{{%
    \setbox0\hbox{\circ}%
    \rlap{\hbox to \wd0{\hss\sum\hss}}\box0
}}

\begin{document}

\title{
Novel Quantum Criticality in Two Dimensional Topological Phase transitions
}   
\author{Gil Young Cho and Eun-Gook Moon}         
\affiliation{Department of Physics, Korea Advanced Institute of Science and Technology, Daejeon 305-701, Korea}
\date{\today}    

\begin{abstract}
Topological quantum phase transitions intrinsically intertwine self-similarity and topology of many-electron  wave-functions, and divining them is one of the most significant ways to advance understanding in condensed matter physics. 
Our focus is {to investigate} an unconventional class of the transitions between insulators and Dirac semimetals whose description is beyond conventional pseudo relativistic Dirac Hamiltonian. 
At the transition without the long-range Coulomb interaction, the electronic energy dispersion along one direction behaves like a relativistic particle, linear in momentum, but along the other direction it behaves like a non-relativistic particle, quadratic in momentum. 
Various physical systems ranging from TiO$_2$-VO$_2$ heterostructure to organic material $\alpha-$(BEDT-TTF)$_2$I$_3$ under pressure have been proposed to have such anisotropic dispersion relation.
Here, we discover a novel quantum criticality at the phase transition by incorporating the $\frac{1}{r}$ long range Coulomb interaction. Unique interplay between the  Coulomb interaction and electronic critical modes enforces not only {the anisotropic renormalization of the Coulomb interaction but also marginally modified electronic excitation.} In connection with experiments, we investigate several striking effects in physical observables of our novel criticality.
\end{abstract}
\maketitle
\nopagebreak
Quantum criticality and topology are two of the main impetuses of modern condensed matter physics.   
Self-similarity of many-electron wave-functions associated with quantum criticality \cite{sachdev, millis, hertz} unveils emergent universality of physical observables, and { topology of the electronic wave-functions manifests itself as various fascinating topological insulators and associated quantized responses\cite{Hasan2011, KaneRev, QiRev, SenthilReview}.}  The two striking characteristics of the wave-function are naturally and inevitably intertwined at topological quantum phase transitions.

Long-range $\frac{1}{r}$ Coulomb interaction between electrons induces striking screening effects near the topological phase transitions. 
Electronic critical modes and the Coulomb interaction are intrinsically correlated 
, so non-trivial quantum criticality usually appears \cite{moon1,savary,yang,hong,lai2015, amaricci2015}. For example, quasi-particles lose their stability due to the Coulomb interaction and the ground state becomes quantum critical non-Fermi liquid with emergent full rotational symmetry {in quadratic band touching semimetals, which is} near three dimensional (3d) topological insulator \cite{moon1}. 

In two dimensions (2d), the Coulomb interaction becomes more special. It is because the Coulomb potential originally lives in 3d but electrons are confined in 2d. {Thus the electrons in 2d feels the dimensionally different interaction, originating from 3d}. Since correlation and fluctuation are enhanced in lower dimensions, one {may} expect stronger interplay between the Coulomb interaction and  critical modes in a topological phase transition, and indeed we find the novel quantum criticality in a class of 2d topological quantum phase transitions.

Conventional 2d topological phase transitions between two topologically distinct insulators are described by the pseudo-relativistic Dirac fermion theory   
$
H^D(\bm{k}) = v_x k_x \sigma^x   + v_y k_y \sigma^y + M \sigma^z, \nonumber
$
with Pauli matrices in band index spinor space.
{ Here the topological nature of the transition is captured by the change in the Berry curvature of the wave-function depending on the sign of $M$,} and  different patterns of opening up band gaps at separate Dirac  points represent different topological insulator phases when supplemented with proper symmetries. The long-range Coulomb interaction at the critical point ($M=0$) induces intriguing logarithmic modification of the Dirac velocities, so not only rotational symmetry at the critical point emerges but also important interaction effects appear whose structure has been extensively studied in  literature \cite{son, sheehy} in connection with charge-neutral mono-layer graphenes. 
We emphasize that the isotropic $\frac{1}{r}$ Coulomb interaction  dominates microscopic anisotropy of electrons in this case.

Here we focus on a different class of the topological phase transitions whose electronic Hamiltonian is 
\begin{align}
H(\bm{k}) = v k_x \sigma^x + A k_y^2\sigma^y.  
\label{1:fermionaction}
\end{align}
With a tuning parameter $m$, energy spectrum of $H = H(\bm{k})- m \sigma^y$ is $E_{\pm}(\bm{k}) = \pm \sqrt{v^2 k_x^2 + (A k_y^2 - m)^2}$. The two phases are determined by the sign of the tuning parameter. With a positve $A$ and $m<0$, energy spectrum is gapped, so the ground state is an insulator. {On the other hand, with $m>0$, the zero-energy points appear at two points in momentum space, $k_y = \pm k_y^{*} = \pm \sqrt{\tfrac{m}{A}}$.} By expanding the Hamiltonian \eqref{1:fermionaction} near these points, we obtain 
$
H_{+}(\bm{p}) = v p_x \sigma^x + v_y  p_y \sigma^y \nonumber
$ 
around the point ${\bf k} = (p_x, k_y^*+p_y)$ with $v_y =2 A k_y^{*}$.
Thus the ground state is a 2d Dirac semimetal.
Thus, it is clear that our model Hamiltonian describes  a phase transition between a (either topological or trivial) insulator and a Dirac semimetal in 2d. 

The Hamiltonian \eqref{1:fermionaction} has been suggested in various physical systems, ranging from TiO$_{2}$-VO$_{2}$ oxide heterostructures\cite{pickett1,pickett2,pickett3} and the organic material $\alpha-$(BEDT-TTF)$_2$I$_3$ under pressure\cite{a1,a2,a3} to optical lattice systems\cite{o1,o2,o3}. 
For example, in the oxide heterostructure TiO$_{2}$-VO$_{2}$ layers\cite{pickett1,pickett2,pickett3}, there is a metal-insulator transition as the number of layers is changed. At the certain number of layers, the first-principle band structure calculation\cite{pickett1,pickett2,pickett3} reveals that there should be the anisotropic semimetal \eqref{1:fermionaction}. Furthermore, the structure of the Hamiltonian \eqref{1:fermionaction} is similar to that of the notorious quantum criticality problem  with Fermi surfaces in 2d \cite{lee1, max1,max2,mross1,nayak1, nayak2} whose scaling of the dispersion along the radial direction to the Fermi surface is linear in momentum while that along the perpendicular direction is quadratic in momentum. Due to the similarity in dispersions, we expect that our analysis might shed some light on understanding the quantum criticality with Fermi surfaces despite of the finite density of states in the Fermi surface case.

In this work, we show, by using the systematic renormalization group (RG) method, that the long range interaction strongly changes the nature of the eigenstates of the non-interacting Hamiltonian. 
We find a novel quantum criticality characterized by {\it both} anisotropically renormalized and marginal Coulomb interactions which is in sharp contrast to other quantum criticalities. The anisotropic marginal quantum criticality is out of intricate interplay between the long range Coulomb interaction in 2d and the critical electron modes, and we emphasize its striking properties by calculating physical observables.

\section*{MODELS WITH COULOMB INTERACTION}
We start with the theory incorporating the electron Hamiltonian with the long-range $\frac{1}{r}$ Coulomb interaction,  
\begin{eqnarray}
\mathcal{S} &=& \mathcal{S}_{\psi} + \mathcal{S}_{\phi} \label{1:bareaction} \\ 
\mathcal{S}_{\psi} &= &\int d^2 x d\tau {\psi^{\dagger}}( (\partial_\tau + ie \phi) + H(-i \nabla) ) \psi \nonumber\\ 
\mathcal{S}_{\phi} &=& \int d^2 x d\tau ~\frac{1}{2} \phi ~\sqrt{|\nabla^2|} ~\phi = \int_{{\bf q}, \omega} \frac{1}{2}|\bm{q}| |\phi(\bm{q},\omega)|^2, \nonumber
\end{eqnarray}
where $\phi$ mediates Coulomb interaction between electrons $\psi$. The short-hand writing $\int_{\bm{q}, \omega} = \int \frac{d^{2}\bm{q}d\omega}{(2\pi)^3}$ is used. 
Hereafter, all integrations are defined with the short-distance  (or high-energy) ultra-violet (UV) cutoff. $H(-i \nabla)$ is the Fourier transformation of $H({\bm k})$ \eqref{1:fermionaction}, and {the bare gauge boson propagator $g_{b,0}^{-1}(\bm{q}) = |\bm{q}|$ represents the long-range $\tfrac{1}{r}$ Coulomb interaction.} 
For future convenience, we introduce a dimensionless coupling constant, the fine structure constant $\alpha = \frac{e^2}{2\pi^2 v}$, which measures the ``strength'' of Coulomb interaction.  

\begin{figure}
\includegraphics[width=6in]{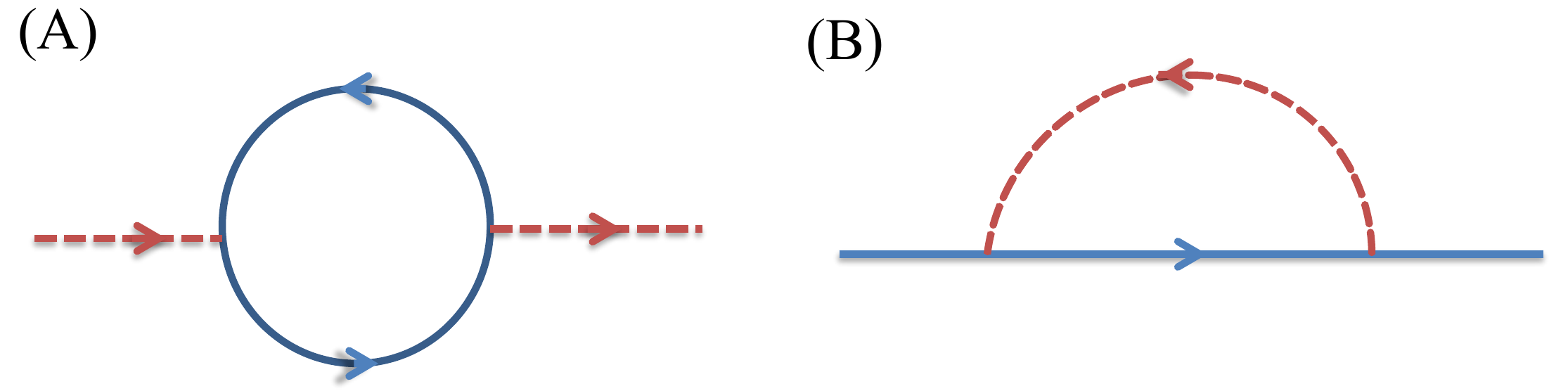}
\caption{
Diagrams for (a) the boson self-energy $\Pi(\bm{q})$ and (b) the fermion self-energy $\Sigma_f(\bm{k}, \omega)$. Here the dotted line represents the boson propagator $g_{b}(\bm{q},\Omega)$ and the solid line represents the fermion propagator $g_{f}(\bm{k}, \omega)$. 
 \label{Diagrams}
}
\end{figure}

We investigate the stability of the theory by the lowest order perturbation calculation, in particular, by calculating the bosonic self-energy whose Feynman diagram representation is in Fig. \ref{Diagrams} (a),
\begin{align}
\Pi(\bm{q}, \Omega) = e^2 \int_{\bm{k}, \omega} \text{Tr}\Big[ g_f(\bm{k}+\bm{q}, \omega+\Omega) g_f (\bm{k}, \omega) \Big]. \nonumber
\end{align}
$g_f^{-1}(\bm{k}, \omega) = -i \omega + H(\bm{k}) $ is used.  It is straightforward to evaluate the integral (see the supplementary information I for detail), and we find that
\begin{align}
\Pi(\bm{q}) = -\frac{\alpha}{2} |q_y| ~ G(\xi_{\bm{q}}), ~~ \xi_{\bm{q}} = \sqrt{\frac{A q_y^2}{v |q_x|}},  
\end{align}
where $G(\xi_{\bm{q}})$ is the function of the \textit{dimensionless parameter} $\xi_{\bm{q}}$. Hereafter, we drop the frequency dependence in the boson self-energy since we are only interested in the instantaneous Coulomb interaction. The full functional form of $G(\xi)$ is not important. Thus we will not present it here and plot it only in the supplementary information I. Instead, the asymptotic behavior of $\Pi(\bm{q})$ in each direction is extracted
\begin{align}
&\Pi(q_x, 0) = -  \frac{\alpha  c_x}{2}  \sqrt{\frac{v|q_x|}{A}},\quad \quad \Pi(0, q_y) = - \frac{\alpha c_y}{2} \times |q_y| \nonumber  
\label{2:polarizationasymptotic}
\end{align}
with the numeric constants $\{ c_x \approx 2.7, c_y \approx 2.5 \}$. 
Notice that the boson self-energy is independent of the UV cutoff which signal a novel quantum criticality in our system.

It is clear that the perturbation becomes more important than the original bare term 
along $\bm{q} = q_x \hat{x}$, 
\begin{align}
|g_{b,0}^{-1} (q_x,0)|  \sim |q_x| \quad \ll \quad |\Pi(q_x,0)| \sim \sqrt{|q_x|},   \nonumber
\end{align}
in the limit $\bm{q} \to \bm{0}$. 
Thus we conclude that the action \eqref{1:bareaction} is {\it unstable} under the fermion-gauge boson coupling. 


The instability from the perturbative calculation often indicates the presence of the stable strong-coupling fixed point which can be accessed by large-$N_f$ analysis with the number of fermion flavors $N_f$. 
The large-$N_f$ analysis starts with adding the bosonic self-energy to the boson bare term,
\begin{align}
\mathcal{S}_{\phi} \,\, \rightarrow \,\, \int_{{\bf q}, \omega} \frac{1}{2}(|\bm{q}|- N_f \Pi(\bm{q}) |\phi(\bm{q},\omega)|^2. \nonumber
\end{align} 
The schematic representation of the  inverse of the corrected boson propagator $g_{b}^{-1} ({\bm q}) = |{\bm q}|- N_f \Pi(\bm{q})$  is
\begin{eqnarray}
g_{b}^{-1} ({\bm q})
&\sim& |{\bm q}|+   \frac{N_f \alpha}{2} \big( c_y|q_y|+   c_x \sqrt{\frac{v |q_x|}{A}}  \big). 
\label{1:boson_N} 
\end{eqnarray}
The limit $\alpha \rightarrow 0$ recovers the unstable bare action \eqref{1:bareaction}, and we investigate the opposite limit $N_f \alpha \rightarrow \infty$ where we drop the bare term ($\sim |{\bf q}|$).

{Using this corrected boson propagator, we calculate} the fermion self-energy in Fig. \ref{Diagrams} (b) 
\begin{align}
-\Sigma_f (\bm{k}, \omega) 
= \delta v( {\Lambda},{\mu}) k_x\sigma^x +\delta A({\Lambda},{\mu}) k_y^2 \sigma^y, \nonumber 
\end{align}
obtained by expanding the self-energy near $\bm{k}=0$ with the UV and IR cutoffs, $\Lambda, \mu$. Straightforward calculation gives
\begin{eqnarray}
\frac{\delta v}{v} = \frac{2 J_x}{N_f} \log(\frac{\Lambda}{\mu}),\quad \frac{\delta A}{A} = \frac{2 J_y}{N_f} \log(\frac{\Lambda}{\mu}), \nonumber 
\label{jxjy}
\end{eqnarray} 
with the two dimensionless constants, $J_x \approx 0.18, J_y  \approx 0.03$ (see supplementary information III for detail). We notice that the instantaneous nature of the Coulomb propagator enforces no vertex correction through the Ward identity.

{Therefore, the RG flow equations, i.e., beta functions, for $v$ and $\alpha$ can be derived by changing the ratio, $\frac{\Lambda}{\mu}=e^{l}$,} 
\begin{align}
\frac{d v} {d l} = \frac{2J_x}{N_f} v, \quad \frac{d \alpha}{dl} = - \frac{2J_x}{N_f} \alpha,  
\end{align}
near the strong-coupling fixed point. It is clear from the RG equations that the fine structure constant $\alpha$ decreases with the anomalous dimension of the velocity, $\frac{2J_x}{N_f}$.  
This concludes that the strong-coupling fixed point is {\it unstable}.  

Our controlled analysis near the two extreme limits (standard perturbation and large-$N_f$ analysis) clearly shows that both the fixed points are unstable.
Then it is obvious that the stable fixed point should be in the intermediate regime, which is difficult to access  in a fully controlled way. Thus, we study the fixed point with the   standard momentum-shell RG and check \textit{{a posteriori}} its validity by self-consistency.   


In the momentum-shell RG analysis,  
we remark that the non-analytic dependence $|{\bf q}|$ of the Coulomb interaction does not receive correction from integrating out higher-momentum modes. Thus, we first keep the seemingly irrelevant $\sim \tilde{\kappa} q_x^2$ term in the boson action,
 \begin{align}
\mathcal{S}_{\phi} \,\,
\rightarrow \,\, \int_{q,\omega} ~\frac{1}{2}(|{\bm q}| + \tilde{\kappa} q_x^2) |\phi(\bm{q},\omega)|^2. \nonumber
\label{BareRG}
\end{align} 
It turns out that the following three dimensionless parameters determine the RG flows
\begin{align}
\alpha = \frac{e^2 }{2\pi^2 v},~~~~ \gamma = \frac{\tilde{\kappa} A^2 \Lambda^3 }{v}, ~~~~ \beta = \frac{\alpha}{3\gamma}. 
\end{align}

Evaluating Feynman diagrams in Fig. \ref{Diagrams} gives
the renormalized action $\mathcal{S}' = \mathcal{S}'_\psi + \mathcal{S}'_\phi$. 
Here, we use the cutoff scheme such that we integrate along the $\hat{y}$-directional momentum ($\Lambda e^{-l}, \Lambda$) with $l \ll 1$ after integrating out the $\hat{x}$-direction momentum and frequency.  
On integrating out the higher-momentum modes, three parameters are renormalized as
\begin{align}
\frac{\delta v}{v} = \alpha F_1[\gamma]\, l, \quad \frac{\delta A}{A} = \alpha F_2[\gamma] \,l, \quad \frac{\delta {\tilde \kappa}}{{\tilde \kappa}} = \beta \, l \nonumber
\end{align}
We find that the functions $F_{1,2}[\gamma]$ are non-negative near $\gamma = 0$ whose specific forms are illustrated in supplementary information IV.
The RG flow equations of $\{ \alpha, \gamma, \beta \}$ are 
\begin{align}
&\frac{d\alpha}{dl} = -\alpha^2 F_1[\gamma],\nonumber\\ 
&\frac{d\gamma}{dl} =  \gamma \Big(\beta-3 + 2\alpha (F_2[\gamma]-F_1[\gamma])\Big),\nonumber\\ 
&\frac{d \beta}{dl} = \beta\Big(3-\beta +  \alpha  (F_1[\gamma]-2 F_2 [\gamma])\Big).
\end{align}

The two fixed points are, $(\alpha, \gamma, \beta) =(0,0,0), (0,0,3)$, and it is easy to show that the former is unstable and the latter is stable. At the stable fixed point $(\alpha, \gamma, \beta) = (0,0,3)$, the boson propagator receives a large anomalous scaling dimension, which can be understood as  $(1+ \beta l)\Big|_{\beta=3}q_x^2 \approx q_x^2 e^{3l} \to \sqrt{|q_x|}$ (because $e^{-2l}$ is the scaling factor of $q_x$). Such large anomalous dimension indicates that the momentum-shell RG is not controlled and {\it a priori} not reliable. 

At the stable fixed point, the effective bosonic action becomes 
\begin{align}
&\mathcal{S}^f_\phi = \int_{\bm{k}, \omega} \frac{1}{2} \Big( |\bm{q}|+ \kappa \sqrt{|q_x|} \Big) |\phi(\bm{q},\omega)|^2
\label{2:finalRG}
\end{align}
which is very similar to the large-$N_f$ calculation with one important difference; a new coupling constant ($\kappa$) with UV cutoff scale  naturally enters in contrast to the large-$N_f$ calculation where the coefficient of $\sqrt{|q_x|}$ is $\sim  \sqrt{\tfrac{v}{A}}$ (see equation \eqref{1:boson_N}) which depends on the other parameters $\{ v, A, \alpha \}$. { Here the new dimensionful parameter $\kappa$ appears in the bosonic part at the intermediate coupling regime}.

With this intuition in hand, we investigate the stability of the new fixed point by taking equation \eqref{2:finalRG} as the bare boson action and performing the momentum-shell RG near this fixed point. Remarkably, we find that the velocity $v$ and inverse mass $A$ receives the same corrections at the fixed point
\begin{align}
& \frac{\delta v}{v}  = \alpha \cdot \mathcal{C}_f  \log(\frac{\Lambda}{\mu}), \quad \frac{\delta A}{A} = \alpha \cdot \mathcal{C}_f \log(\frac{\Lambda}{\mu}), 
 \nonumber
 \end{align}
The same correction is another evidence for our fixed point to be stable since the ratio $\sqrt{\tfrac{v}{A}}$ {  appearing in the boson self-energy $\Pi(\bm{q})$} becomes constant. Notice that the remarkable same correction also appears in Fermi surface quantum criticalities with very different physical reasons which also supports stability of our fixed point\cite{max1}. { It is manifest that the gapless excitation structure of our system is completely different from that of Fermi surfaces (lines) in 2d. Only nodal point excitation appears in our system. However, the low energy scaling structures of the two systems are same considering the patch theory of Fermi surfaces in 2d ; one momentum direction has linear scaling while the other one has quadratic scaling. We believe this unexpected similarity is the source of the similar behaviors in the beta functions. It would be very intriguing to find more similarity and difference of two systems' quantum criticalities, which we leave for future work.  }

The beta functions around the novel fixed point are 
\begin{align}  
& \frac{d \alpha}{dl} = - C_f \alpha^2, \quad  \frac{d v}{d l } = \alpha \cdot C_f v, \quad \frac{d A}{d l } = \alpha \cdot C_f A 
\label{2:scaling}
\end{align}
where $\mathcal{C}_f \approx 0.8$ calculated in the supplementary information V. 

{  We remark that the hard momentum cutoff scheme is only used for simplicity and illustration. It is shown that our results are independent of cutoff schemes in the supplementary information VIII.}

The above RG flow structure \eqref{2:scaling} is unique to this fixed point. The beta functions contain the fine structure constant $\alpha$ in contrast to those of the large-$N_f$ calculation \eqref{jxjy} in which $\alpha$ is absent, and here both $v, A$  receive the {\it same} logarithmic corrections, which are proportional to $\alpha$. Thus, the fine structure constant $\alpha$ decreases and the fermion only receives the logarithmic corrections, which indicates the fixed point is {\it stable}. Naturally, as in mono-layer graphene,  marginal Fermi liquid behaviors are expected with higher order corrections.\cite{gonzalez}

\begin{figure}
\begin{center}
\includegraphics[width=6.0in]{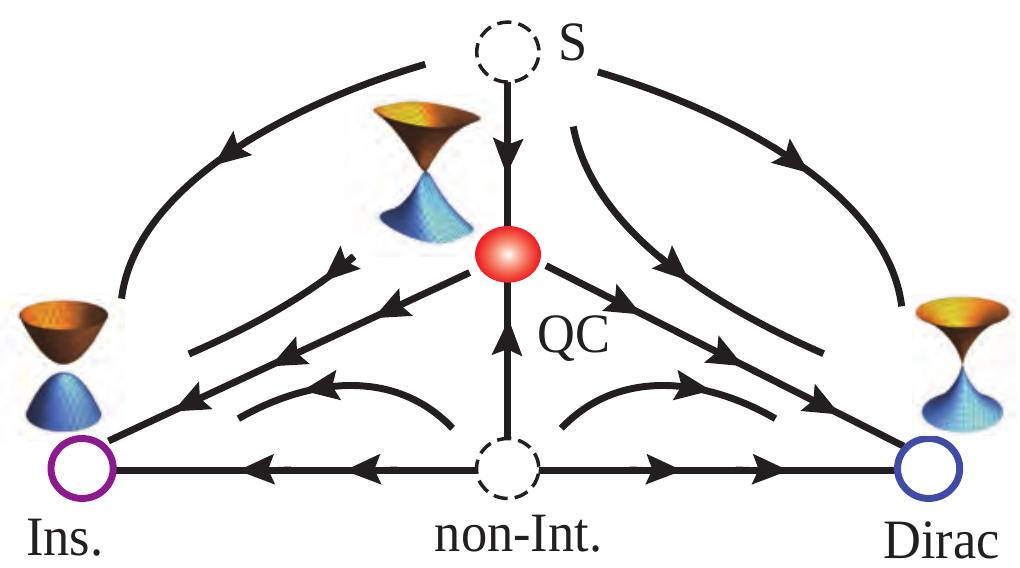}
\caption{
Proposed RG flow. The horizontal axis is for the tuning parameter $m$ of the quantum criticality equation \eqref{1:fermionaction} and the vertical axis is for the strength of Coulomb interaction. 
There are two stable fixed points, insulators (`Ins') and Dirac semimetal (`Dirac'). The two unstable critical points are illustrated with dashed circles, non-interacting (`Non-Int.') and strong-coupling fixed point (`S'). And the stable critical point is the filled circle (`QC'). The critical point is characterized by the definite anisotropic scaling and the logarithmic corrections to mass and velocity. Near the fixed points (Ins, Dirac, QC), one-particle spectrum with the Coulomb interaction is illustrated. 
 \label{RGflow}
}
\end{center}
\end{figure}

Based on the calculations and intuitions, A schematic RG flow can be deduced as in Fig. \ref{RGflow} which summarizes our main results. 
Our controlled calculation shows the non-interacting critical point (Non-Int.) and the strong coupled fixed point (S) are unstable and the RG flow comes out of the both points and flow into the intermediate fixed point (QC), which is characterized by the definite anisotropic scaling of bosons and electrons and the single logarithmic corrections to velocity and inverse mass.

\section*{EXPERIMENTAL SIGNATURES}\label{experimentalsignatures}
We now investigate the physical consequences of  both the anisotropy and marginal irrelevance of the renormalized Coulomb potential at the novel intermediate critical point. 

First of all, with the beta functions of $v$ and $A$ \eqref{2:scaling}, we can find logarithmic corrections to all physical quantities. The parameters $\{v, A \}$ at the temperature scale $T$ are  
\begin{align}
v(T) = v_0 (1+ \alpha C_f \log (\frac{E_0}{T})), \quad
A(T) = A_0 (1+ \alpha C_f  \log (\frac{E_0}{T})).
\label{Escaling}
\end{align}
Here $E_0$ is the bandwidth or the UV cutoff of the theory \eqref{1:fermionaction}. $v_0$ and $A_0$ are the bare parameters at the highest energy scale $\sim E_0$. The logarithmic corrections in $\{v, A \}$ may be observed in quasi-particle experiments as in graphene, for example, angle resolved photo-emission spectroscopy (ARPES)\cite{arpes} or quantum oscillation. 

Furthermore, thermodynamic quantities such as specific heat and compressibility also show logarithmic corrections. 
Specific heat and compressibility of the unstable free electron fixed point are 
$C_v(T) = \frac{\partial  E(T) }{\partial T} \approx  \frac{0.38 \,T^{3/2}}{v_0 \sqrt{A_0}}$ and $\kappa(T)= \frac{\partial  n(\mu, T) }{\partial \mu} \Big|_{\mu \to 0} \approx  \frac{0.07\,\sqrt{T}}{v_0 \sqrt{A_0}}$
in which $E(T)$ is the (thermal-averaged) energy density per volume as the function of temperature $T$ and $n(\mu, T)$ is the density of the electron per volume as the function of chemical potential $\mu$ and temperature $T$. But at the novel fixed point, the logarithmic corrections give  
\begin{align}
C_v(T) \approx \frac{T^{3/2}}{v_0 \sqrt{A_0}} \frac{0.38}{(1+ \alpha C_f  \log(\frac{E_0}{T}))^{3/2}}, \quad
\kappa (T) \approx \frac{0.07 \sqrt{T}}{v_0 \sqrt{A_0} (1+\alpha C_f  \log(\frac{E_0}{T}))^{3/2} }, 
\end{align}
at the temperature $T$ by following the reference\cite{sheehy}. 

Secondly, we can see the effect of the anisotropic renormalization of the gauge boson via the screening charge when a single impurity charge $Z$ is introduced at $\bm{r} =\bm{0}$. At the level of the linear response theory, the screening charge is 
$
\rho_{\text{ind}} (\bm{q}) = Z D(\bm{q}) \Pi_0 (\bm{q})
$,
in which $D(\bm{q})$ is the propagator of the gauge boson. We are interested in the \textit{directional} behaviors of the screening charge and hence define the integrated screening charges $Q_x(x)= \int^{\infty}_{-\infty} dy  \rho_{\text{ind}} (\bm{r})$ and $Q_y(y) =  \int^{\infty}_{-\infty} dx  \rho_{\text{ind}} (\bm{r})$ along $\hat{x}$ and $\hat{y}$. Here we will contrast the extremely different behaviors of the screening charges between the free fixed point and the non-trivial fixed point.

At the free fixed point, we ignore the corrections to the gauge boson propagator and use $D_0 (\bm{q}) =\frac{1}{|\bm{q}|}$. Following the straightforward calculation in supplementary information VI, we find $Q_x(x) \propto -\frac{Z}{\sqrt{|x|}}, ~ Q_y(y) \propto -Z\delta(y)$ whose sign is the opposite of the impurity charge $Z$. 

On the other hand, at the non-trivial anisotropic fixed point in which we use the renormalized boson propagator $D_{\text{rem}} (\bm{q})^{-1} = |\bm{q}| + \kappa \sqrt{|q_x|}$, we find  the asymptotic behaviors of the screening charges   
\begin{align}
Q_x (x)  \propto \frac{Z}{|x|(1+ \alpha C_f  \log (|x/r_0|))^2}, \quad
Q_y(y) \propto  \frac{Z}{|y|(1+ \alpha C_f  \log (|y/r_0|))^2}, 
\end{align}
where $r_0^{-1} \sim E_0$ is the UV cutoff. Here the sign of the screening charge is the same as the impurity charge $Z$, which is reminiscent of graphene case\cite{biswas}. 

From the above calculations, we see that the asymptotic scaling behaviors of the screening charges in distance from the impurity along $\hat{x}$ and $\hat{y}$ are surprisingly isotropic. The isotropic scaling in both the directions is originated from the facts that the scaling of $D_{\text{rem}}(\bm{q})$ is identical to that of $\Pi_{0} (\bm{q})$ and that $\{v(\mu), A(\mu)\}$ at the energy scale $\mu$ receive the same logarithmic corrections as \eqref{2:scaling}. Hence this isotropic scaling behaviors are truly from the effects of interactions between the electrons and the gauge boson.

\begin{table}[tb!]
\begin{tabular} {c|c|c}
\hline
Systems & Excitation & Coulomb  \\ \hline \hline
2D Dirac\cite{son,gonzalez}  & marginal q.p.& iso., marginally irr. \\ \hline
3D Dirac\cite{hosur}  & marginal q.p. & iso., marginally irr. \\ \hline
3D Quadratic\cite{moon1}   & no q.p. & iso., relevant  \\ \hline
3D Anisotropic\cite{yang} & q.p. &  aniso., irr.  \\ \hline
2D Anisotropic  & marginal q.p.  &  aniso., marginal \\ \hline
\end{tabular}
\caption{Comparison with the quantum criticalities in various semimetallic systems. Here the second column represents types of allowed excitation. ``q.p.'' is for quasi-particle. The third column represents characteristics of screened Coulomb interaction. ``iso.'' is for isotropic, ``aniso.`` is for anisotropic, and ``irr.'' is for irrelevant.  }
\label{tab:result}
\end{table}

\section*{DISCUSSION AND CONCLUSION}\label{conclusion}
The presence of the novel fixed point implies that the electrons and gauge bosons are strongly correlated. At low energy, electrons and gauge bosons affect each other, so the Coulomb interaction mediated by the bosons becomes anisotropic and electrons receive back-reaction from the renormalized anisotropic Coulomb interaction. 
Thus, the Coulomb interaction behaves differently from that of most critical systems where it enforces low-energy isotropy of electronic modes\cite{hosur,son,gonzalez, moon1}.  
Also, notice that the ground state of our fixed point has marginally well-defined quasi-particles as those in graphene, which is in contrast to non-Fermi liquids with non-zero anomalous dimensions. In table \ref{tab:result}, the comparison with other quantum criticality associated with topological phase transitions is summarized.

Our novel quantum criticality can be experimentally tested in the systems such as VO$_2$-TiO$_2$ heterostructure. Near the critical point, optical conductivity shows anisotropy inherited from the electron band structure. Straightforward calculation with current operators $(j_x, j_y) = (\psi^{\dagger} \sigma^x \psi, \psi^{\dagger} {A k_y} 
 \sigma^y \psi)$  gives  
\begin{align}
\sigma_{xx} (\Omega) \propto \frac{1}{\sqrt{\Omega}}, \quad \sigma_{yy} (\Omega) \propto \sqrt{\Omega}, 
\end{align}
upto logarithmic corrections from the Coulomb interaction (see the supplementary information VII). However, as shown in the previous section, the screening charge due to the charged impurity, which can be measured in principle by scanning tunneling microscopy (STM), shows qualitatively isotropic behaviors. Such discrepancy between the two experiments is a smoking gun of the novel quantum criticality in addition to thermodynamics quantities such as specific heat. 

It is worth to mention that disorder scattering in the non-interacting electrons \eqref{1:fermionaction} is relevant\cite{carpentier}, so our results work better for cleaner samples. We expect that there will be an intriguing interplay between the anisotropic Coulomb interaction and impurity scattering at the novel critical point, which we leave for the future problem.

In conclusion, we have investigated the quantum criticality of the anisotropic semimetal which can be thought as the critical point between topological insulators and Dirac semimetal in two spatial dimensions. At the low-energy limit, we found the novel  fixed point out of the interplay between critical electron modes and the long-range $\frac{1}{r}$ Coulomb interaction. 
The non-trivial anisotropic renormalization of the Coulomb interaction and the logarithmic corrections manifest at various physical quantities including screening charge when the impurity charge is introduced. Surprisingly we have shown that the scaling behavior of the screening charge in distance from the impurity is isotropic despite of the underlying anisotropic nature of the system.  

\textit{-Note added}: After the completion of the paper, we became aware of the independent work by H. Isobe, B.-J. Yang, A. Chubukov, J. Schmalian, and N. Nagaosa [\onlinecite{Isobe2015}]. Similarity and differences between our work and theirs
are discussed in supplementary information.

\textit{-Author Contributions}: G.Y.C. performed most calculation and E.-G. M. came up with the original idea. Both authors analyzed data and wrote the manuscript. 

\textit{-Competing Financial Interests}: The authors declare no competing financial interests.

\begin{acknowledgments}
We thank Yong Baek Kim and Charles Kane for useful discussions. We are grateful to Naoto Nagaosa and Andrey Chubukov for sending their manuscript and having discussions.   
This work was supported by the Brain Korea 21 PLUS Project of Korea Government and KAIST start-up funding. 
\end{acknowledgments}



\newpage
\begin{widetext}
\section*{Supplementary Information for ``Novel Quantum Criticality in Two Dimensional Topological Phase transitions"}
\end{widetext}
\setcounter{section}{0}

\section{Polarization Bubble}\label{polarizationbubble}
In this supplemental method, we present the detailed calculation of the polarization, or the boson self-energy $\Pi(\bm{k})$ appearing in the main text. For convenience, we assume $N$-copies of the electrons coupled to the gauge boson. If we are interested in the only one copy, i.e., physical limit, one can simply take $N=1$ at the end of the calculation. 

\subsection{Deivation of Polarization}
In the leading order in $1/N$ expansion, the boson self-energy is simply the two-leg one-loop diagram. The fermion propagator appearing in the loop diagram is 
\begin{align}
g_f (\bm{k}, \omega) = \frac{1}{-i\omega + H(\bm{k})}, 
\end{align}
where $H(\bm{k})$ is the free fermion Hamiltonian equation (6) in the main text. It is convenient to write the propagator in the following form \begin{align}
g_f (\bm{k}, \omega) &= \frac{1}{-i\omega + H(\bm{k})} =\sum_{s = \pm} \frac{1}{-i\omega + E_s (\bm{k})} P_s (\bm{k}), 
\end{align}
with the energy $E_s (\bm k) = s E(\bm{k}) = s \sqrt{v^2 k_x^2 + A^2 k_y^4}$ and the projection $P_s (\bm{k})$ 
\begin{align}
P_s (\bm{k}) = \frac{1}{2} \Big(1+ s \frac{H(\bm{k})}{E(\bm{k})}\Big). 
\end{align}
Now the polarization is given by 
\begin{align}
\Pi(\bm{p}, \Omega) = \frac{e^2}{N} \int_{\bm{k}, \omega} \text{Tr}\Big[ g_f(\bm{k}+\frac{\bm{p}}{2}, \omega+\Omega) g_f (\bm{k} -\frac{\bm{p}}{2}, \omega) \Big]. 
\end{align}
Performing the contour integral over $\omega$ and the trace $\text{Tr} [ \cdot ]$, we finally have 
\begin{align}
\Pi(\bm{p}, \omega) = -e^2 N \int_{\bm{k}} \frac{\mathcal{P}(\bm{k}, \bm{p}) \left(E(\bm{k} + \frac{\bm{p}}{2}) + E(\bm{k} - \frac{\bm{p}}{2})\right)}{\omega^2 + \left(E(\bm{k} + \frac{\bm{p}}{2}) + E(\bm{k} - \frac{\bm{p}}{2})\right)^2},
\label{Pol}
\end{align}
in which $\mathcal{P}(\bm{k}, \bm{p})$ is the projection operator 
\begin{align}
\mathcal{P}(\bm{k}, \bm{p}) = 1- \frac{\sum_{\mu = 1,2}\epsilon_{\mu}(\bm{k} + \frac{\bm{p}}{2}) \epsilon_{\mu} (\bm{k} - \frac{\bm{p}}{2})}{E(\bm{k} + \frac{\bm{p}}{2}) E(\bm{k} - \frac{\bm{p}}{2})}, 
\end{align}
where $\epsilon_{1}(\bm{k}) = v k_x$ and $\epsilon_2 (\bm{k}) = A k_y^2$ are the appropriate functions.
 
\subsection{Calculation of Polarization}
\begin{figure}
\begin{center}
\includegraphics[width=6.0in]{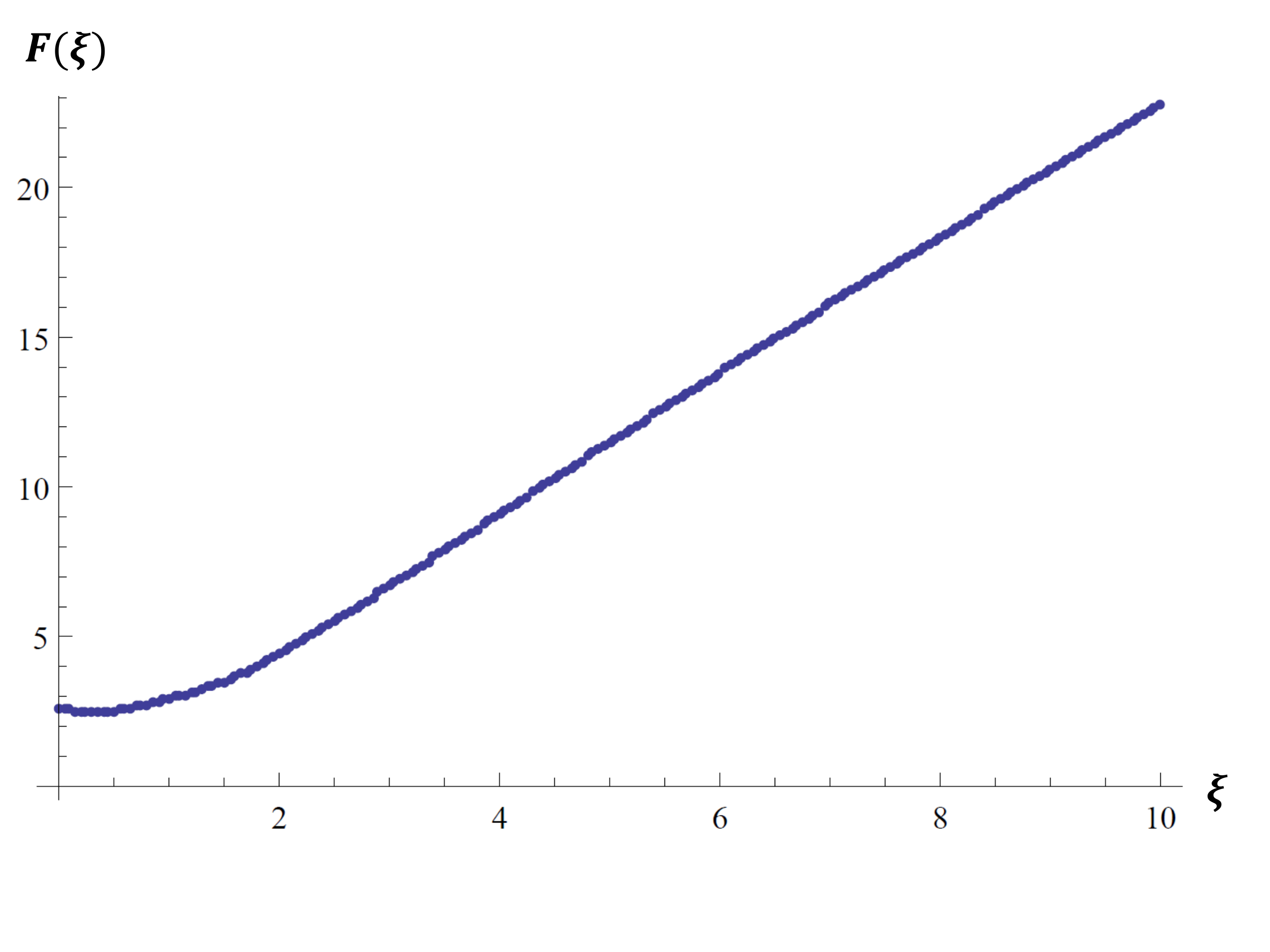}
\caption{
Plot of $F(\xi)$ 
 \label{Fxi}
}
\end{center}
\end{figure}

To evaluate the function $\Pi(\bm{k}) \equiv \Pi(\bm{k}, \omega = 0)$ in equation\eqref{Pol}, we use the change of variables as following: 
\begin{align}
k_x = p_x x, ~\text{ and }~ k_y = \sqrt{\frac{v|p_x|}{A}} y,  
\end{align}
which gives rise to the following expression. 
\begin{align}
\Pi(\bm{p}) = -\frac{e^2 N}{4\pi^2 v} \sqrt{\frac{v|p_x|}{A}} \int^{\infty}_{-\infty} dx \int^{\infty}_{-\infty} dy ~f_{\xi} (x,y), 
\end{align}
where $\xi = |p_y| \sqrt{\frac{v|p_x|}{A}}$ and$ f_{\xi}(x,y)$ is the function 
\begin{align}
f_{\xi}(x,y) = \frac{\Big[1- \frac{x^2 -\frac{1}{4} + (y^2 - \frac{\xi^2}{4})^2}{\sqrt{(x+\frac{1}{2})^2 + (y + \frac{\xi}{2})^4}\sqrt{(x-\frac{1}{2})^2 + (y - \frac{\xi}{2})^4}}\Big]}{\sqrt{(x+\frac{1}{2})^2 + (y + \frac{\xi}{2})^4}+\sqrt{(x-\frac{1}{2})^2 + (y - \frac{\xi}{2})^4}}.  
\end{align}
We introduce a function 
\begin{align}
F(\xi) = \int^{\infty}_{-\infty} dx \int^{\infty}_{-\infty} dy ~f_{\xi} (x,y). 
\end{align}
Then it is easy to see $F(\xi) = F(-\xi)$, i.e., $F(\xi) = F(|\xi|)$. Now, we can see that 
\begin{align}
F(|\xi| \to 0) \to c_x, ~\text{ and }~ F(|\xi| \to \infty) \to c_y \xi,   
\label{asym}
\end{align}
where $c_x \approx 2.7$ and $c_y \approx 2.5$. We plot $F(\xi)$ in the Fig. \ref{Fxi}. From the numerical calculation, we can extract both the constants. With these in hand, we can write the polarization as 
\begin{align}
\Pi(\bm{p}) &= -\frac{e^2 N}{4\pi^2 v}|p_y| \frac{F(|\xi|)}{|\xi|} =  -\frac{e^2 N}{4\pi^2 v}|p_y| G(|\xi|)\nonumber\\ 
&=-\frac{\alpha N}{2 }|p_y| G(|\xi|), 
\end{align}
where we have used $G(|\xi|) = \frac{F(|\xi|)}{|\xi|}$ and $\alpha = \frac{e^2}{2\pi^2 v}$. From equation\eqref{asym}, we can see that 
\begin{align}
G(|\xi| \to 0) \to \frac{c_x}{\xi}, ~\text{ and }~ G(|\xi| \to \infty) \to c_y,  
\end{align}
and $G(|\xi|) >0 $ for any $\xi$. 

\section{Screening effects in quantum criticalities}
Here, we summarize bare and self-energy of Coulomb boson propagators in various quantum criticalities. 
The one-loop boson self-energy is obtained in the standard way. 
With different electron Hamiltonian for each case, we obtain the self-energy which is illustrated in Table \ref{tab:self}.

\begin{table}[tb!]
\begin{tabular} {c|c|c}
\hline
Systems & Bare & Boson Self-energy  \\ \hline \hline
2D Dirac\cite{son,gonzalez}  & $|{\bm q}|$ & $|{\bm q}|$ \\ \hline
3D Dirac\cite{hosur}  & ${\bm q}^2$ & ${\bm q}^2 \log \Lambda$ \\ \hline
3D Quadratic\cite{moon1}   & ${\bm q}^2$ & ${|\bm q|}$  \\ \hline
3D Double Weyl\cite{hong, lai2015}  & ${\bm q}^2$ & ${ q}_{\perp}^2 \log \Lambda + |q_z|$ \\ \hline
3D Anisotropic\cite{yang} & ${\bm q}^2$  &  $\Lambda|{ q}_{\perp}|^2+|q_z|^{\frac{3}{2}}$  \\ \hline
2D Anisotropic  & $|{\bm q}|$  & $ \sqrt{|q_x|} + |q_y|$ \\ \hline
\end{tabular}
\caption{Bare and self-energy of boson propagators in various  quantum criticalities. The second and thir column represent  the bare and the one-loop self-energy of various quantum criticalities. $\Lambda$ is the UV cutoff and only schematic behaviors are shown. Note that 3D quantum criticalities depend on the UV cutoff except the non-Fermi liquid quadratic band touching. }
\label{tab:self}
\end{table}

\section{Fermion Self-Energy and RG equations in Strong-Coupling Limit}\label{strongsupplemental method}
Here we compute the fermion self-energy at the large-N limit to extract the RG equations. We start with the expression for the fermion self-energy with the corrected boson propagator 
\begin{align}
\Sigma_f (\bm{k}, \omega) = -\frac{e^2}{2} \int_{\bm{q}} \frac{H(\bm{q}+\bm{k})}{E(\bm{q}+\bm{k})} \frac{1}{|\bm{q}| + \frac{\alpha N}{2} |q_y|G(\xi_{\bm{q}})}. 
\end{align}
To extract out the correction to the bare parameters $\{ v, A \}$, we expand the $\Sigma_f (\bm{k}, \omega)$ in $\bm{k}$ near $\bm{k}=0$, 
\begin{align}
-\Sigma_f (\bm{k}) = v k_x \sigma^x \tilde{I}_x + A k_y^2 \sigma^y \tilde{I}_y. 
\label{selfF}
\end{align} 
Here $\{ \tilde{I}_x, \tilde{I}_y \}$ are the integrals 
\begin{align}
&\tilde{I}_x = \frac{e^2}{8\pi^2} \int dq_x dq_y \frac{A^2 q_y^4}{[v^2 q_x^2 + A^2 q_y^4]^{3/2}}\frac{1}{|\bm{q}| + \frac{\alpha N}{2} |q_y|G(\xi_{\bm{q}})}, \nonumber\\ 
&\tilde{I}_y = \frac{e^2}{8\pi^2} \int dq_x dq_y \frac{v^4 q_x^4- 5 v^2 q_x^2 A^2 q_y^4}{[v^2 q_x^2 + A^2 q_y^4]^{5/2}}\frac{1}{|\bm{q}| + \frac{\alpha N}{2} |q_y|G(\xi_{\bm{q}})}. 
\label{Int}
\end{align}
Apparently, the integrals are divergent and needed to be regulated. We regulate the integrals by 
\begin{align}
\int dq_x dq_y \to \Big(\int^{\Lambda}_{\mu} dq_x + \int^{-\mu}_{-\Lambda} dq_x\Big) \int^{\infty}_{-\infty} dq_y,
\label{cut}
\end{align}
where the UV and IR cutoffs $\{ \Lambda, \mu \}$ are introduced. 

With the expression equation\eqref{selfF}, we find 
\begin{align}
\delta v &= -\frac{1}{\text{Tr}[\sigma^0]}\text{Tr} \Big[ \sigma^x \frac{\delta \Sigma_f (\bm{k}, \omega)}{\delta ( k_x)} \Big]_{\bm{k} \to \bm{0}, \omega \to 0} \nonumber\\ 
&= v \tilde{I}_x \nonumber\\ 
\delta A &= -\frac{1}{\text{Tr}[\sigma^0]}\text{Tr} \Big[ \sigma^y \frac{\delta \Sigma_f (\bm{k}, \omega)}{\delta ( k_y^2)} \Big]_{\bm{k} \to \bm{0}, \omega \to 0}\nonumber\\
&= A \tilde{I}_y
\label{relation}
\end{align} 

\subsection{Strong-coupling limit}
We first evaluate the integrals equation\eqref{Int} when $\alpha \to \infty$. 
\begin{align}
&\tilde{I}_x = \frac{e^2}{4\pi^2 \alpha N } \int dq_x \int dq_y \frac{A^2 q_y^4}{[v^2 q_x^2 + A^2 q_y^4]^{3/2}} \frac{1}{|q_y| G(\xi_{\bm{q}})},\nonumber\\ 
&\tilde{I}_y = \frac{e^2}{4\pi^2 \alpha N} \int dq_x \int dq_y  \frac{v^4 q_x^4- 5 v^2 q_x^2 A^2 q_y^4}{[v^2 q_x^2 + A^2 q_y^4]^{5/2}}\frac{1}{|q_y| G(\xi_{\bm{q}})}, 
\end{align}
in which the cutoff scheme equation\eqref{cut} is assumed. We demonstrate the detail of evaluations only for $\tilde{I}_x$.  
\begin{align}
\tilde{I}_x &= \frac{e^2}{4\pi^2 \alpha N} \int dq_x \int dq_y \frac{A^2 q_y^4}{[v^2 q_x^2 + A^2 q_y^4]^{3/2}} \frac{1}{|q_y| G(\xi_{\bm{q}})},\nonumber\\
&=  \frac{e^2}{4 \pi^2 \alpha N v} \int \frac{dq_x}{q_x} \int \frac{dq_y}{|q_y|} \frac{\frac{A^2 q_y^4}{v^2q_x^2}}{[1 + \frac{A^2 q_y^4}{v^2 q_x^2}]^{3/2}} \frac{1}{G(\xi_{\bm{q}})}\nonumber\\ 
&= \frac{1}{2N} \int \frac{dq_x}{q_x} \int^{\infty}_{-\infty} d\xi \frac{\xi^3}{(1+\xi^4)^{3/2}}\frac{1}{G(\xi)}\nonumber\\
&=\frac{2}{N} \log (\frac{\Lambda}{\mu}) \int^{\infty}_{0} d\xi \frac{\xi^3}{(1+\xi^4)^{3/2}} \frac{1}{G(\xi)} \nonumber\\ 
&=\frac{2J_x}{N} \log (\frac{\Lambda}{\mu}) 
\end{align}
where 
\begin{align}
J_x = \int^{\infty}_{0} d\xi \frac{\xi^3}{(1+\xi^4)^{3/2}} \frac{1}{G(\xi)}. 
\end{align}
The integral is well-defined in that it has no divergence in the integral because 
\begin{align}
&\frac{\xi^3}{(1+\xi^4)^{3/2}} \frac{1}{G(\xi)}\Big|_{\xi \to \infty}  \to\frac{1}{\xi^3} \frac{1}{c_x} \propto \frac{1}{\xi^3}\nonumber\\
&\frac{\xi^3}{(1+\xi^4)^{3/2}} \frac{1}{G(\xi)}\Big|_{\xi \to 0}  \to \xi^3 \frac{1}{c_y/\xi} \propto \xi^4
\end{align}
The integral can be evaluated numerically by using the numeric value of $G(\xi)$, and we have obtained $J_x \approx 0.18$. The numeric value is not important but it is important to remember $J_x >0$ because of $G(|\xi|) >0$. 

Similarly, we can write $\tilde{I}_y$ as following. 
\begin{align}
\tilde{I}_y = \frac{2 J_y}{N} \log(\frac{\Lambda}{\mu}), 
\label{reny}
\end{align}
where $J_y$ is the definite integral 
\begin{align}
J_y = \int^{\infty}_0 \frac{d\xi}{\xi} \frac{1-5 \xi^4}{(1+ \xi^4)^{5/2}}\frac{1}{G(\xi)}. 
\end{align}
The integeral is well-defined because 
\begin{align}
&\frac{1-5 \xi^4}{(1+ \xi^4)^{5/2}}\frac{1}{\xi G(\xi)}\Big|_{\xi \to \infty}  \to\frac{-5}{\xi^7} \frac{1}{c_y} \propto \frac{1}{\xi^7}\nonumber\\ 
&\frac{1-5 \xi^4}{(1+ \xi^4)^{5/2}}\frac{1}{\xi G(\xi)}\Big|_{\xi \to 0}  \to\frac{1}{\xi} \frac{1}{c_x/\xi} \propto \frac{1}{c_x}.
\end{align}
Hence there is no more IR divergence than $\log (\Lambda/\mu)$ appearing in equation\eqref{reny} in the renormalization of $A$ in the strong-coupling limit. The numeric value of $J_y$ is obtained as $0.03$. 

Hence, under the RG step, we find that the velocity $v$ and the inverse mass $A$ are renormalized as follows. 
\begin{align}
&v \to v + \delta v  = v (1+ \frac{2J_x}{N} \log (\Lambda)),\nonumber\\ 
&A \to A +\delta A = A (1+ \frac{2J_y}{N} \log (\Lambda)), 
\end{align}
from which we deduce the RG flow equations, 
\begin{align}
&\frac{d v} {d l} = \frac{2J_x}{N} v,\nonumber\\ 
&\frac{d A} {d l} = \frac{2J_y}{N} A, 
\end{align}
which have been shown in the main text. 

Notice that the frequency dependence is trivial since only the instantaneous Coulomb propagator is used. If one incorporates frequency dependence, only the numeric numbers $J_x, J_y$ are modified but not the structure of our calculation. The absence of the frequency dependence indicates no vertex correction by the Ward identity. 

\subsection{Finite Coupling Constant}
Here we will discuss the fate of the finite coupling constant $\alpha$ which is now slightly away from the strong-coupling limit. In the prsence of the finite coupling constant $\alpha$, the bare term $\sim |\bm{q}|$ in the boson propagator 
\begin{align}
g_{b}(\bm{q},\omega)= \frac{1}{|\bm{q}| - \Pi(\bm{q})}
\end{align} 
cannot be ignored. Hence the above calculations of $\alpha \to \infty$ should be properly changed to see the scaling behaviors of the parameters $\{ v, A \}$.   

Performing the straightforward calculation as the above, we find 
\begin{align}
\frac{\delta v}{v} = \alpha K_x (\Lambda, \mu),~~ \frac{\delta A}{A} = \alpha K_y (\Lambda, \mu)
\end{align}
The integrals $\{ K_x, K_y \}$ are 
\begin{align}
&K_x = \int^{\Lambda}_{\mu} \frac{dq_x}{q_x} \int^{\infty}_0 d\xi  \frac{\frac{\xi^4}{(1+\xi^4)^{3/2}}}{(\xi^2+ \bar{A}(q_x))^{1/2} + \frac{\alpha N}{2} \xi G(\xi)},  \nonumber\\ 
&K_y = \int^{\Lambda}_{\mu} \frac{dq_x}{q_x} \int^{\infty}_0 d\xi \frac{\frac{1-5\xi^4}{(1+\xi^4)^{5/2}}}{(\xi^2+ \bar{A}(q_x))^{1/2} + \frac{\alpha N}{2} \xi G(\xi)}, 
\label{integral}
\end{align}
where $\bar{A}(q_x) = \frac{A q_x}{v}$ is the dimensionless number depending on the momentum $q_x$.  

Before marching further into the details, we first make an assumption, $\bar{A}(q_x) < \bar{A}(\Lambda) \ll 1$. The condition $\bar{A}(\Lambda) \ll 1$ should be interpreted as following. First of all, $v\Lambda$ can be roughly thought as the bandwidth where the low-energy Hamiltonian equation (6) in the main text is the qualitatively correct description of the system. Note that $A$ determines the curvature along $k_y$ and $v$ determines the velocity along $k_x$. Thus $\bar{A}(\Lambda) \ll 1$ means that the velocity along $k_y$ is always much smaller than the velocity along $k_x$ within the energy $E \leq v \Lambda$, i.e., the dispersion along the $k_y$ direction is always much flatter than that along the $k_x$ direction. In other words, the spectrum is \textit{extremely anisotropic} in that the equal energy contour is extended along $k_y$ and shrinked along $k_x$. 

At this stage, the assumption $\bar{A}(\Lambda) \ll 1$ looks like an ad-hoc assumption to evaluate the integrals equation\eqref{integral}. However, employeeing the momentum-shell RG approach at the weak-coupling limit $\alpha \to 0$ (as present in the main text and the next supplemental method), we can show $\bar{A} \ll 1$ at the fixed point.

With this condition in hand, we proceed to evaluate the integrals equation\eqref{integral} approximately by taking $\bar{A} \to 0$, 
\begin{align}
&K_x = \log (\frac{\Lambda}{\mu}) \tilde{J}_x (\alpha) \nonumber\\ 
&K_y = \log(\frac{\Lambda}{\mu}) \tilde{J}_y  (\alpha)
\end{align}
where $\tilde{J}_x >0$ and $\tilde{J}_y$ are the constants depending on $\alpha$ whose detailed calculation can be found below. From these, we find 
\begin{align}
&\frac{\delta v}{v} = \alpha \log (\frac{\Lambda}{\mu}) \tilde{J}_x (\alpha), \nonumber\\
&\frac{\delta A}{A} = \alpha \log(\frac{\Lambda}{\mu}) \tilde{J}_y (\alpha).  
\end{align}
Hence we find that 
\begin{align}
&\frac{\delta \log(v)}{\delta \log(\Lambda)} = \alpha \tilde{J}_x (\alpha)\nonumber\\ 
&\frac{\delta \log(A)}{\delta \log(\Lambda)} = \alpha \tilde{J}_y (\alpha).
\end{align}
Thus the parameters $\{ v, A\}$ in the bare fermion Hamiltonian equation (6) in the main text receive the logarithmic corrections, and the coupling constant $\alpha$ decreases along the renormalization group process.  

\subsubsection{Calculation of $\tilde{J}_x$ and $\tilde{J}_y$}
Below we only present the detail for evaluating $\tilde{J}_x$ but it is straightforward to generalize to $\tilde{J}_y$. 
\begin{align}
&\tilde{J}_x = \frac{e^2}{8\pi^2} \int dq_x dq_y \frac{A^2 q_y^4}{[v^2 q_x^2 + A^2 q_y^4]^{3/2}}\frac{1}{|\bm{q}| + \frac{\alpha N}{2} |q_y|G(\xi_{\bm{q}})}, \nonumber\\ 
&= \frac{e^2}{4\pi^2 v} \int^{\Lambda}_{\mu} \frac{dq_x}{q_x} \int^{\infty}_{-\infty} dq_y \frac{\frac{A^2q_y^4}{v^2 q_x^2}}{(1+ \frac{A^2 q_y^4}{v^2 q_x^2})^{3/2}} \frac{1}{|\bm{q}| + \frac{\alpha N}{2} |q_y|G(\xi_{\bm{q}})}\nonumber\\ 
&=\frac{\alpha}{2} \int^{\Lambda}_{\mu} \frac{dq_x}{q_x} \int^{\infty}_{-\infty} \frac{dq_y}{|q_y|} \frac{\xi^4}{(1+\xi^4)^{3/2}}\frac{1}{\sqrt{1+ \frac{q_x^2}{q_y^2}} + \frac{\alpha N}{2} G(\xi)}\nonumber\\
&=\alpha \int^{\Lambda}_{\mu} \frac{dq_x}{q_x} \int^{\infty}_{0} \frac{d\xi}{\xi} \frac{\xi^4}{(1+\xi^4)^{3/2}}\frac{1}{\sqrt{1+ \frac{\bar{A}(q_x)}{\xi^2}} + \frac{\alpha N}{2} G(\xi)},
\end{align}
in which $\bar{A}(q_x) = \frac{A}{v} q_x$ and $q_y = \sqrt{\frac{vq_x}{A}} \xi$. We continue to evaluate this 
\begin{align}
&\tilde{J}_x =\alpha \int^{\Lambda}_{\mu} \frac{dq_x}{q_x} \int^{\infty}_{0} \frac{d\xi}{\xi} \frac{\xi^4}{(1+\xi^4)^{3/2}}\frac{1}{\sqrt{1+ \frac{\bar{A}(q_x)}{\xi^2}} + \frac{\alpha N}{2} G(\xi)} \nonumber\\ 
&=\alpha \int^{\Lambda}_{\mu} \frac{dq_x}{q_x} \int^{\infty}_{0}d\xi \frac{\xi^4}{(1+\xi^4)^{3/2}}\frac{1}{\sqrt{\xi^2+ \bar{A}(q_x)} + \frac{\alpha N}{2} \xi G(\xi)}
\end{align}
Now assuming $\bar{A}(q_x) < \bar{A}(\Lambda) \ll 1$ which we will justify in the low-energy limit later, we can ignore the dependence on $\bar{A}$ in the integral to obtain 
\begin{align}
&\tilde{J}_x \approx \alpha \int^{\Lambda}_{\mu} \frac{dq_x}{q_x} \int^{\infty}_{0}d\xi \frac{\xi^4}{(1+\xi^4)^{3/2}}\frac{1}{\xi + \frac{\alpha N}{2} \xi G(\xi)}\nonumber\\ 
&= \alpha \int^{\Lambda}_{\mu} \frac{dq_x}{q_x} \int^{\infty}_{0}d\xi \frac{\xi^3}{(1+\xi^4)^{3/2}}\frac{1}{1 + \frac{\alpha N}{2} G(\xi)}\nonumber\\
&= \alpha \log(\frac{\Lambda}{\mu})\tilde{J}_x (\alpha)
\end{align}
where 
\begin{align}
\tilde{J}_x (\alpha) = \int^{\infty}_{0}d\xi \frac{\xi^3}{(1+\xi^4)^{3/2}}\frac{1}{1 + \frac{\alpha N}{2} G(\xi_{\bm{q}})}. 
\end{align}
This integral is well-defined and finite as 
\begin{align}
&\frac{\xi^3}{(1+\xi^4)^{3/2}}\frac{1}{1 + \frac{\alpha N}{2} G(\xi_{\bm{q}})}\Big|_{\xi \to \infty} \to \frac{1}{\xi^3} \nonumber\\ 
&\frac{\xi^3}{(1+\xi^4)^{3/2}}\frac{1}{1 + \frac{\alpha N}{2} G(\xi_{\bm{q}})}\Big|_{\xi \to 0} \to \xi^4. 
\end{align}
Also $\tilde{J}_x(\alpha)$ is always positive as $G(\xi)>0$.  

On the other hand, we can evaluate $\tilde{J}_y$ similarly in $\bar{A} \ll 1$ and obtain 
\begin{align}
\tilde{J}_y = \alpha \log(\frac{\Lambda}{\mu}) \tilde{J}_y(\alpha),
\label{reny2} 
\end{align}
in which 
\begin{align}
\tilde{J}_y(\alpha) = \int^{\infty}_{0}\frac{d\xi}{\xi} \frac{1-5 \xi^4}{(1+\xi^4)^{5/2}} \frac{1}{1+ \frac{\alpha N}{2} G(\xi)}. 
\end{align}
The integral is finite and well-defined as 
\begin{align}
&\frac{1-5 \xi^4}{(1+\xi^4)^{5/2}} \frac{1}{1+ \frac{\alpha N}{2} G(\xi)} \Big|_{\xi \to \infty} \propto \frac{1}{\xi^7} \nonumber\\ 
&\frac{1-5 \xi^4}{(1+\xi^4)^{5/2}} \frac{1}{1+ \frac{\alpha N}{2} G(\xi)} \Big|_{\xi \to 0} \propto \frac{1}{c_x}. 
\end{align}
Hence there is no more divergence in the renormalization in the inverse mass than $\propto \log(\frac{\Lambda}{\mu})$ in equation\eqref{reny2}. 

\section{Detailed Calculation of Momentum Shell RG}\label{momentumshell}
In this supplemental method, we will perform the one-loop RG by the momentum-shell method. We start with the following bare action $\mathcal{S} = \mathcal{S}_\psi + \mathcal{S}_\phi + \mathcal{S}_{\psi,\phi}$ present in the main text,   
\begin{align}
&\mathcal{S}_\psi = \int_{\bm{k}, \omega} \psi^{\dagger}_{\bm{k},\omega} (-i\omega + v k_x \sigma^x + A k_y^2 \sigma^y)\psi_{\bm{k}, \omega}, \nonumber\\   
&\mathcal{S}_\phi = \int_{\bm{k}, \omega} \frac{1}{2} \Big(\sqrt{\eta q_x^2  + \frac{q_y^2}{\eta}} + \tilde{\kappa} q_x^2 \Big) |\phi(\bm{q},\omega)|^2,  \nonumber\\ 
&\mathcal{S}_{\psi,\phi} = \int_{\bm{k}, \omega} \int_{\bm{q}, \Omega} i e ~\phi(\bm{q}, \omega) \psi^{\dagger}_{\bm{k}+\bm{q}/2, \Omega+\omega} \psi_{\bm{k}-\bm{q}/2, \Omega}, 
\label{BareRG}
\end{align}
where $\eta$ is introduced to facilitate the engineering dimensional analysis of the anisotropic scaling along $x$ and $y$ in the bare boson propagator. The physical value of $\eta$, as mentioned in the main text, is $1$ because the (bare) Coulomb interaction is isotropic.  On the other hand, we have introduced $\kappa$ into the bare action to allow the possible anomalous screening term $\sim \sqrt{|q_x|}$, via the anomalous scaling dimension appearing in $\sim q_x^2$ term, under the RG steps. 

We start with the dimensional analysis. First of all, we set the engineering dimensions $[x]= -z_1, [y] = -1$ and $[\tau] = - z_2$. Then it is straightforward to show  
\begin{align}
&[v] = z-z_1, ~[A] = z-2, ~ [e^2] = z- \frac{1+z_2}{2}, \nonumber\\ 
&[\eta] = 1-z_1, ~[\Lambda] = 1, ~[\tilde{\kappa}] = \frac{1-3 z_1}{2}
\end{align}
where $\Lambda$ is the momentum cutoff in the momentum $q_y$.  Here the fields have the following dimensions. 
\begin{align}
[\phi] = \frac{1}{2}\Big(z+ \frac{1+z_1}{2}\Big), ~[\psi] = \frac{1+z_1}{2} 
\end{align}
On the other hand, in terms of physical unit, we have 
\begin{align}
&[v] = \frac{L}{T}, ~[A] = \frac{L^2}{T}, ~ [e^2] = \frac{L}{T}, \nonumber\\ 
&[\eta] = 1, ~[\Lambda] = \frac{1}{L}, ~[\kappa] = L
\end{align}
The field operators have the following scaling dimensions. 
\begin{align}
[\phi] = \frac{1}{\sqrt{LT}}, ~[\psi] = \frac{1}{\sqrt{L}}
\end{align}
From the above dimensional analysis we have the following dimensionless constants, which will be used to parametrize the renormalization flow
\begin{align}
\alpha = \frac{e^2 \sqrt{\eta}}{2\pi^2 v}, ~ \bar{A} = \frac{A \eta \Lambda}{v}, ~ \gamma = \frac{\tilde{\kappa} A^2 \Lambda^3 \sqrt{\eta}}{v}. 
\end{align}

Before proceeding to the details of the RG calculation, we first note that $\eta$ and $e$ are protected under the RG process due to the non-analytic structure of the propagator of the boson and the gauge invariance. Hence, the only non-trivial corrections are the renormalizations of $\{v, A \}$ in the fermion propagator and $\kappa$ in the boson propagator. The renormalization can be deduced from the self-energies of the fermion and the boson, which we calculate below. 

We start with the fermion self-energy. 
\begin{align}
-\Sigma_f (\bm{k}, \omega) = v k_x \sigma^x I_1 + A k_y^2 \sigma^y I_2,
\end{align}
such that 
\begin{align}
&I_1 = \frac{e^2}{2} \int_{\bm{q}} \frac{1}{\sqrt{\eta q_x^2  + \frac{q_y^2}{\eta}} + \tilde{\kappa} q_x^2} \frac{A q_y^4}{\Big[v^2 q_x^2 + A^2 q_y^4 \Big]^{3/2}},\nonumber\\ 
&I_2 = \frac{e^2}{2} \int_{\bm{q}} \frac{1}{\sqrt{\eta q_x^2  + \frac{q_y^2}{\eta}} + \tilde{\kappa} q_x^2} \frac{v^2 q_x^2(v^2q_x^2-5 A^2 q_y^4)}{\Big[ v^2 q_x^2 + A^2 q_y^4 \Big]^{5/2}},
\label{FerSelf}
\end{align}
where the integrals should be properly regularized as following. 
\begin{align}
\int_{\bm{q}} &= \frac{1}{4\pi^2} \Big( \int^{\Lambda}_{\mu} dq_y + \int^{-\mu}_{-\Lambda} dq_y \Big) \int^{\infty}_{-\infty} dq_x = \frac{1}{\pi^2} \int^{\Lambda}_{\mu} dq_y \int^{\infty}_{0} dq_x, 
\label{Regularization}
\end{align}
as the integrands of equation\eqref{FerSelf} are even under $q_i \to -q_i, i =x,y$. Here $\mu = \Lambda e^{-l} \approx \Lambda (1-l)$ with $l \ll 1$ is assumed by following the standard strategy of the momentum-shell calculations. 
  
With the assumption $\bar{A} \ll 1$ which will be justified self-consistently in the end of the calculation, we can evalute the integrals straightforwardly as following. 
\begin{align}
&I_1 = \alpha l F_1 [\gamma], \nonumber\\ 
&I_2 = \alpha l F_2 [\gamma], 
\end{align}
where 
\begin{align}
&F_1 [\gamma] = \int^{\infty}_{0} dx \frac{1}{1+ \gamma x^2} \frac{1}{(x^2+1)^{3/2}}, \nonumber\\  
&F_2[\gamma] = \int^{\infty}_0 dx \frac{1}{1+\gamma x^2} \frac{x^2 (x^2 -5)}{(x^2+1)^{5/2}}.
\end{align}
We remark that in $\gamma \rightarrow 0$ limit ($\tilde{\kappa} \rightarrow 0$), the inverse mass correction $F_2$ diverges logarithmically in $\sim \log \gamma$. This indicates instability of the non-interacting fermion ground state with the bare Coulomb interaction without operator insertions.   

We now compute the polarization by the momentum-shell integral method 
\begin{align}
-\Pi(\bm{p}) = p_x^2 I_b,
\end{align}
in which 
\begin{align}
I_b &= \frac{e^2}{4\pi^2} \int^{\Lambda}_\mu dk_y \int^{\infty}_0 dk_x \frac{v^2 A^2 k_y^4}{[v^2 k_x^2+A^2 k_y^4]^{5/2}}=\frac{1}{3} \frac{\alpha}{A^2 \Lambda^3 \sqrt{\eta}/v^2} l.  
\end{align}

Now we have the renormalized action $\mathcal{S}' = \mathcal{S}'_\psi + \mathcal{S}'_\phi + \mathcal{S}'_{\phi, \psi}$obtained by integrating out the high-energy modes as the following. We start with the fermion part $\mathcal{S}'_\psi$ where 
\begin{align}
&\mathcal{S}'_\psi = \int d^3 x ~\bar{\psi}(\partial_0 + i e \phi) \psi + \int d^3 x ~\bar{\psi}(v(1+ \alpha F_1[\gamma] l)\gamma_1 \partial_x - A (1+\alpha F_2 [\gamma]l)(\partial_2)^2)\psi. 
\end{align}
On the other hand, the boson part $\mathcal{S}'_\phi$ is  
\begin{align} 
\mathcal{S}'_\phi =  \int d^3 x ~\frac{1}{2}\phi (\sqrt{\eta q_x^2  + \frac{q_y^2}{\eta}} + \tilde{\kappa} (1+ \beta l) q_x^2) \phi, 
\end{align} 
where 
\begin{align}
\beta = \frac{1}{3} \frac{\alpha}{\gamma}. 
\end{align}
On the other hand, the vertex correction vanishes and thus $\mathcal{S}'_{\phi, \psi}$ is the same as $\mathcal{S}_{\phi,\psi}$ from equation\eqref{BareRG}. 

From these, we can obtain the flow equations. 
\begin{align}
&\frac{d}{dl} \log v= z-z_1 + J_1 \nonumber\\ 
&\frac{d}{dl} \log A= z-2 + J_2 \nonumber\\
&\frac{d}{dl} \log \eta = 1 - z_1 \nonumber\\  
&\frac{d}{dl} \log \tilde{\kappa} = \frac{1-3 z_1}{2} + \beta \nonumber\\
&\frac{d}{dl} \log e^2 = z-\frac{1+z_1}{2}  \nonumber\\  
\end{align} 

\begin{figure}
\begin{center}
\includegraphics[width=\columnwidth]{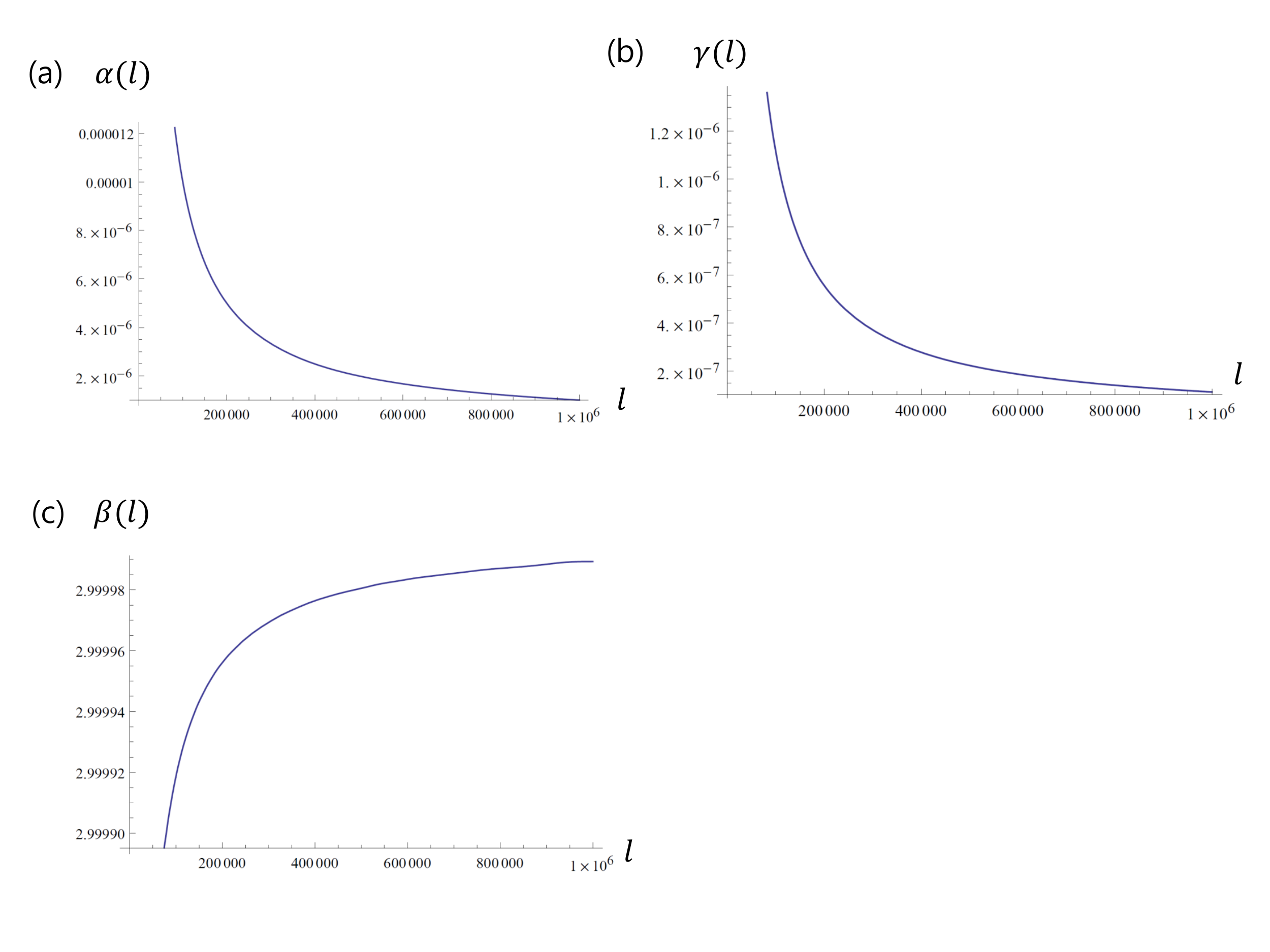}
\caption{
Plot of RG flows. Here $l$ represents the RG time. We have taken the initial conditions $\alpha[0] = \gamma[0] = 0.1$. 
 \label{Flow1}
}
\end{center}
\end{figure}

We now derive the following renormalization flow equations 
\begin{align}
&\frac{d\alpha}{dl} = -\alpha^2 F_1[\gamma],\nonumber\\ 
&\frac{d\gamma}{dl} = \frac{\alpha}{3} + \gamma [-3 + 2\alpha (F_2[\gamma]-F_1[\gamma])],\nonumber\\ 
&\frac{d \beta}{dl} = \beta(3-\beta) + \alpha\beta [F_1[\gamma]-2 F_2 [\gamma]]. 
\end{align}
The fixed point of these equations is $(\alpha, \gamma, \beta) \to (0,0,3)$. We plot the numerical solutions also in Fig.\ref{Flow1}. This also implies that $(\alpha F_1[\gamma], \alpha F_2[\gamma]) \to (0,0)$. With this fixed point, we now derive the fixed point action $\mathcal{S}^{f} = \mathcal{S}^{f}_\psi + \mathcal{S}^{f}_\phi + \mathcal{S}^{f}_{\phi, \psi}$ 
\begin{align}
&\mathcal{S}^f_\psi = \int_{\bm{k}, \omega} \psi^{\dagger}_{\bm{k},\omega} (-i\omega + v k_x \sigma^x + A k_y^2 \sigma^y)\psi_{\bm{k}, \omega}, \nonumber\\   
&\mathcal{S}^f_\phi = \int_{\bm{k}, \omega} \frac{1}{2} \Big( |\bm{q}|+ \kappa \sqrt{|q_x|} \Big) |\phi(\bm{q},\omega)|^2,  \nonumber\\ 
&\mathcal{S}^f_{\psi,\phi} = \int_{\bm{k}, \omega} \int_{\bm{q}, \Omega} i e ~\phi(\bm{q}, \omega) \psi^{\dagger}_{\bm{k}+\bm{q}/2, \Omega+\omega} \psi_{\bm{k}-\bm{q}/2, \Omega}, 
\label{FinalRG1}
\end{align}
where we have used $\eta = 1$ for physical limit and $(1+ \beta l)\Big|_{\beta=3}q_x^2 \approx q_x^2 e^{3l} \to \sqrt{|q_x|}$ because $e^{-2l}$ is the scaling factor of $q_x$.  

Now we justify $\bar{A} \ll 1$ which has been used in the derivations of the fermion self-energy. We calculate 
\begin{align}
\frac{d}{dl} \log \bar{A} = -1 + J_2 - J_1
\end{align}
and we can approximate $J_2 = J_1 \approx 0$ near the fixed point. Then we clearly have 
\begin{align}
\frac{d}{dl} \log \bar{A} = -1 
\end{align}
which dictates that $\bar{A} \to 0 \ll 1$ at the fixed point. 
 
\section{Stability of New Fixed Point}\label{RGNewFixedPoint}
In this supplemental method, we diagnose the stability and the scaling properties near the new intermediate fixed point. To determine the stability of the fixed point and the behavior near the fixed point more carefully, we now take equation\eqref{FinalRG1} as the bare action and perform the momentum shell renormalization analysis again. Taking equation\eqref{FinalRG1} as the bare action, we have 
\begin{align}
[\kappa] = \frac{1}{2} = \sqrt{\frac{1}{L}}. 
\end{align}
Within the theory equation\eqref{FinalRG1}, we can introduce a dimensionless number 
\begin{align}
\bar{\kappa} = \kappa \sqrt{\frac{A\eta}{v}}
\end{align} 
which will determine the renormalization flow near the fixed point equation\eqref{FinalRG1}. Due to the non-analyticity and the gauge invariance, $\{e, \eta, \kappa \}$ are protected under the renormalization process. Then the only non-trivial renormalizations occur in $\{v, A \}$, coming from the fermion self-energy. We again evaluate the fermion self-energy 
\begin{align}
-\Sigma_f (\bm{k}, \omega) = v k_x \sigma^x I'_1 + A k_y^2 \sigma^y I'_2 
\end{align}
in which 
\begin{align}
&I'_1 = \frac{e^2}{2} \int_{\bm{q}}\frac{1}{\sqrt{\eta q_x^2  + \frac{q_y^2}{\eta}} + \kappa \sqrt{|q_x|}}\frac{A^2 q_y^4}{\Big[v^2 q_x^2 + A^2 q_y^4 \Big]^{3/2}} \nonumber\\ 
&I'_2 = \frac{e^2}{2} \int_{\bm{q}}\frac{1}{\sqrt{\eta q_x^2  + \frac{q_y^2}{\eta}} + \kappa \sqrt{|q_x|}}\frac{v^2 q_x^2(v^2q_x^2-5 A^2 q_y^4)}{\Big[ v^2 q_x^2 + A^2 q_y^4 \Big]^{5/2}}
\label{Mom}
\end{align}
where the momentum integral $\int_{\bm{q}}$ is regularized as in equation\eqref{Regularization}. 

After the straightforward algebra with $\bar{A} \ll 1$ near the fixed point, we find 
\begin{align}
&I'_1 = \alpha l f_1[\bar{\kappa}]\nonumber\\ 
&I'_2 = \alpha l f_2[\bar{\kappa}]
\end{align} 
where 
\begin{align}
&f_1[\bar{\kappa}] = \int^{\infty}_0 dx \frac{1}{1+\bar{\kappa} \sqrt{x}} \frac{1}{(x^2+1)^{3/2}}\nonumber\\
&f_2[\bar{\kappa}] = \int^{\infty}_0 dx \frac{1}{1+\bar{\kappa} \sqrt{x}} \frac{x^2(x^2-5)}{(x^2+1)^{5/2}}.
\end{align}

\begin{figure}
\begin{center}
\includegraphics[width=\columnwidth]{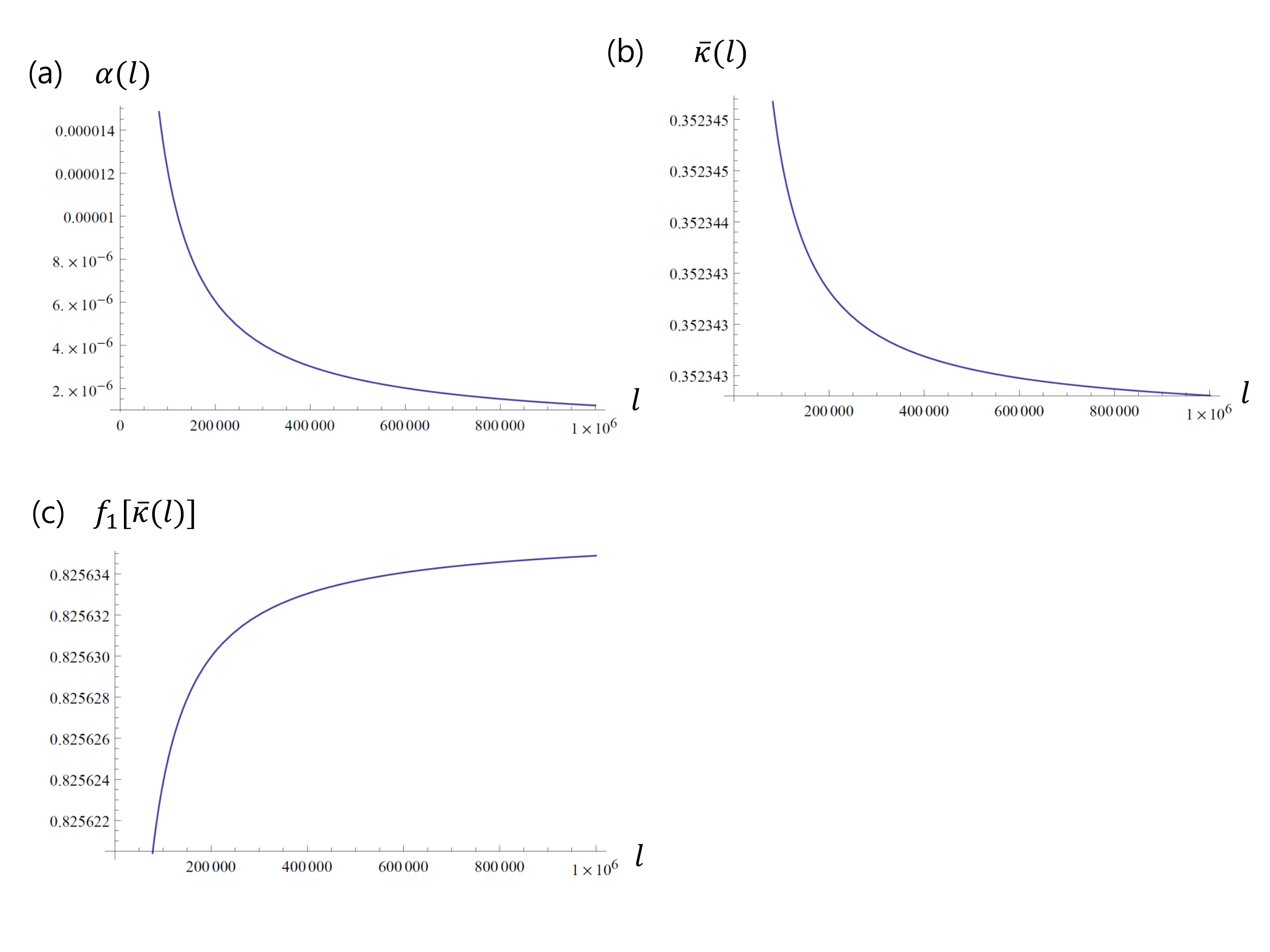}
\caption{
Plot of RG flows. Here $l$ represents the RG time. We have taken the initial conditions $\alpha[0] = \bar{\kappa}[0] = 0.5$. 
 \label{Flow2}
}
\end{center}
\end{figure}

From these, we can derive the flow equations 
\begin{align}
&\frac{d}{dl} \log v= z-z_1 + \alpha f_1[\bar{\kappa}] \nonumber\\ 
&\frac{d}{dl} \log A= z-2 + \alpha f_2[\bar{\kappa}] \nonumber\\
&\frac{d}{dl} \log \eta = 1 - z_1 \nonumber\\  
&\frac{d}{dl} \log \kappa = \frac{1}{2} \nonumber\\
&\frac{d}{dl} \log e^2 = z-\frac{1+z_1}{2}  \nonumber\\  
\end{align} 
which implies 
\begin{align}
&\frac{d}{dl} \alpha = -\alpha^2 f_1[\bar{\kappa}] \nonumber\\ 
&\frac{d}{dl} \bar{\kappa} = \frac{\alpha}{2} \bar{\kappa} [f_2 [\bar{\kappa}] - f_1 [\bar{\kappa}]]. 
\end{align}
The attractive fixed point of these flow equations is $(\alpha, \bar{\kappa}) \to (0, \bar{\kappa}^{*}\approx 0.35)$ such that $f_2[\bar{\kappa}^{*}] = f_1[\bar{\kappa}^{*}]$. The numerical solution can be found in Fig.\ref{Flow2}. Hence we find that the velocity and inverse mass at the energy scale $\mu$ are  
\begin{align}
&v (\mu) \approx v (1+ \alpha C_f \log(\frac{\Lambda}{\mu})) \nonumber\\ 
&A (\mu) \approx A (1+\alpha C_f \log(\frac{\Lambda}{\mu})) 
\label{scaling}
\end{align}
where $\mathcal{C}_f = f_1 [\bar{\kappa}^{*}]\approx 0.8$. This logarithmic divergence as $\mu \to 0$ is reminiscent of graphene case. 

\section{Detailed Calculation of Screening Charge}\label{screening}
In this supplemental method, we show the detailed calculation of the screening charge dictated by equation (18) in the main text and equation (20) in the main text when a single impurity charge $Z$ is introduced at $\bm{r} =0$. We calculate the free and interacting cases at the level of the linear response theory. The details of the calculation here will closely follow Biswas, Son, and Sachdev\cite{biswas}. 

\subsubsection{Free Theory}
We first compute the screening charge of the non-interacting theory from equation (7) in the main text. The screening charge is given by  
\begin{align}
\rho^f_{\text{ind}} (\bm{q}) = Z D_0 (\bm{q}) \Pi_0 (\bm{q})
\end{align}
in which $D_0 (\bm{q}) = \frac{1}{|\bm{q}|}$ and $\Pi_0 (\bm{q})$ is the one-loop polarization. The real space configuration of the screening charge, we perform the Fourier transformation 
\begin{align}
\rho^f_{\text{ind}} (\bm{r}) = \int \frac{d^2 q}{4\pi^2} e^{i \bm{q} \cdot \bm{r}} \rho^f_{\text{ind}} (\bm{q})
\end{align}

We will be mainly interested in the directional behavior of the scaling in the integrated screening charges $Q_x(x)$ and $Q_y(y)$ along $\hat{x}$ and $\hat{y}$, which are obtained from 
\begin{align}
&Q_x (x) = \int^{\infty}_{-\infty} dy  ~\rho^f_{\text{ind}} (\bm{r})\nonumber\\ 
&Q_y (y) = \int^{\infty}_{-\infty} dx  ~\rho^f_{\text{ind}} (\bm{r})
\end{align}

We now evaluate these screening charges. We start with 
\begin{align}
Q_x (x) &= \int^{\infty}_{-\infty} dy  ~\rho^f_{\text{ind}} (\bm{r}) \nonumber\\ 
&= \int \frac{dq_x}{2\pi} \rho^f_{\text{ind}} (q_x, 0) e^{iq_x x} \nonumber\\ 
&= Z \int \frac{dq_x}{2\pi}  D_0 (q_x,0) \Pi_0 (q_x,0) e^{iq_x x} 
\end{align}
Now from 
\begin{align}
\int^x_0 Q_x (x')dx' &\approx \rho^f_{\text{ind}}(q_x = \frac{1}{|x|})\nonumber\\ 
&= Z  D_0 (q_x,0) \Pi_0 (q_x,0) \Big|_{q_x = \frac{1}{|x|}}\nonumber\\ 
&= -Z\frac{e^2}{4\pi^2} \frac{c_x}{\sqrt{vA}} \frac{1}{\sqrt{|q_x|}}\Big|_{q_x = \frac{1}{|x|}}, 
\end{align}
we can take the derivative with respect to $x$ on the both sides. After the straightforward calculation, we have 
\begin{align}
Q_x(x) = -\frac{Z \alpha c_x}{8\pi^2}\sqrt{\frac{v}{A}} \frac{1}{\sqrt{|x|}}
\end{align}
Similarly, we have the following expression for the other direction 
\begin{align}
Q_y (y) = Z \frac{\partial}{\partial y} \Big( D_0 (0,q_y) \Pi_0(0, q_y) \Big)\Big|_{q_y = \frac{1}{|y|}}. 
\end{align}
By plugging 
\begin{align}
D_0 (0,q_y) \Pi_0(0, q_y) = \frac{e^2 c_y}{4\pi^2 v}, 
\end{align}
we find that 
\begin{align}
Q_y(y) \propto \delta (y). 
\end{align}
Hence we see that the free theory has completely different scaling behaviors of the integrated screening charge along $\hat{x}$ and $\hat{y}$ directions, i.e., 
\begin{align}
Q_x(x) \propto -\frac{Z}{\sqrt{x}}, ~ Q_y(y) \propto -Z\delta(y).
\end{align}
whose sign is the opposite of the impurity charge $Z$. 

\subsubsection{Interacting Theory}
Now we use the non-trivial anisotropic fixed point equation (18) in the main text obtained from the interactions between the fermion and the boson. We will use the renormalized boson propagator dictated from equation (18) in the main text and the scaling behavior equation (20) in the main text to deduce the asymptotic behaviors of the integrated screening charges. 
\begin{align}
&\rho^f_{\text{ind}} (\bm{q}) \approx Z D_{\text{rem}} (\bm{q}) \Pi_0 (\bm{q}),\nonumber\\ 
&D_{\text{rem}} (\bm{q})^{-1} = |\bm{q}| + \kappa \sqrt{|q_x|}. 
\end{align}

From this, we now calculate the integrated screening charge $Q_x(x)$. To calculate it, we need  
\begin{align}
\Pi_0(q_x,0) = - \frac{e^2 c_x}{4\pi^2 v(q_x)} \sqrt{\frac{v(q_x)}{A(q_x)}}\sqrt{|q_x|} 
\end{align}
in which $\{v(q), A(q) \}$ are the running parameters at the scale $q$. From equation\eqref{scaling}, we have 
\begin{align}
\Pi_0(q_x,0) = - \frac{e^2 c_x}{4\pi^2 v(q_x)} \sqrt{\frac{v_0}{A_0}}\sqrt{|q_x|}, 
\end{align}
where $\{v_0, A_0\}$ are the bare parameters. Hence we find 
\begin{align}
\rho_{\text{ind}}(q_x) = -Z \frac{e^2 c_x}{4\pi^2 \kappa v(q_x)} \sqrt{\frac{v_0}{A_0}}, 
\end{align}
where we have $v(q_x) = v_0 (1+ \alpha C_f  \log (\frac{1}{r_0 q_x}))$ with the $r_0^{-1} \sim E_0 $, the bandwidth of the bare action equation (6) in the main text. From the relationship 
\begin{align}
\int^x_0 Q_x (x')dx' &\approx \rho_{\text{ind}}(q_x = \frac{1}{|x|}), 
\end{align}
we find 
\begin{align}
Q_x (x) &\approx \frac{Z (\alpha C_f )^2 c_x}{2 f_1 [\bar{\kappa}^{*}]} \sqrt{\frac{v_0}{A_0}} \frac{1}{|x|(1+ \alpha C_f  \log (|x|/r_0))^2}\nonumber\\ 
& \propto \frac{Z}{|x|(1+ \alpha C_f  \log (|x|/r_0))^2}, 
\end{align}
whose sign is the same as the impurity charge $Z$. 

From 
\begin{align}
\Pi(0, q_y) = - \frac{e^2 c_y}{4\pi^2 v(q_y)} |q_y|,  
\end{align}
we find 
\begin{align}
\rho_{\text{ind}}(0, q_y) = -\frac{e^2 Z c_y}{4\pi^2 v(q_y)}. 
\end{align}
Thus we see that 
\begin{align}
Q_y(y) &\approx \frac{(\alpha C_f )^2 Z c_y}{2 f_1[\bar{\kappa}^{*}]} \frac{1}{|y| (1+ \alpha C_f  \log (|y|/r_0))^2}\nonumber\\ 
& \propto  \frac{Z}{|y| (1+ \alpha C_f  \log (|y|/r_0))^2}
\end{align}
whose sign is the same as the impurity charge $Z$. 

\section{Detailed Calculation of Optical Conductivity}\label{optical}
In this supplemental method, we compute the optical conductivity $\sigma_{jj}(\Omega), j = x,y$ at the level of the linear response theory. The expression for the optical conductivity is given by 
\begin{align}
\sigma_{jj}(\Omega)= \frac{1}{\Omega} \text{Im}\Big[ \Pi_{jj}^{R}(\Omega + i\delta)\Big], ~ j=x,y
\label{def_op}
\end{align}
in which $\Pi^{R}_{ii}(\Omega + i \delta) \equiv \Pi_{ii}(i\Omega \to \Omega + i \delta)$ where the latter is obtained from the imaginary-time formalism. Hence we need to compute 
\begin{align}
&\Pi_{jj}(i\Omega) = \int_{\vec{k}, \omega} \text{Tr}\Big[ M_{j} g_{f} (\vec{k}, \omega+\Omega) M_{j} g_f(\vec{k}, \Omega) \Big],  \nonumber\\ 
&=\sum_{\alpha,\beta = \pm1}\int_{\vec{k}} \text{Tr}\Big(M_j P_{\vec{k},\alpha} M_j P_{\vec{k},\beta}\Big)\frac{n_F(\alpha)-n_F(\beta)}{-i\Omega + \alpha E_{\vec{k}}-\beta E_{\vec{k}}}
\end{align}
in which $M_j$ is the appropriate vertex along the $j$-direction and the factor $P_{\vec{k},\alpha} = \frac{1}{2}\Big(1+ \alpha \frac{H_{\vec{k}}}{E_{\vec{k}}}\Big), \alpha = \pm1$ comes from the Green's function $g_{f}(\vec{k},\omega)$. By performing summation over $\{ \alpha, \beta \}$, it is straightforward to find 
\begin{align}
\Pi_{jj}(i\Omega) = \int_{\vec{k}} \text{Tr}&\Big( M_j P_{\vec{k}, +1} M_j P_{\vec{k},-1}\Big) \times  \Big(\frac{-1}{-i\Omega + 2E_{\bm{k}}} + \frac{1}{-i\Omega + 2 E_{\bm{k}}}\Big) 
\end{align}
Here we perform the Wick rotation $i\Omega \to \Omega + i\delta$. 
\begin{align}
\Pi_{jj}^{R}(\Omega) = \int_{\vec{k}} \text{Tr}&\Big( M_j P_{\vec{k},+1} M_j P_{\vec{k},-1}\Big)  \times \Big(\frac{-1}{\Omega + i\delta - 2E_{\bm{k}}} - \frac{1}{\Omega +i\delta + 2 E_{\bm{k}}}\Big) 
\end{align}
Now it is straightforward to take the imaginary part from this expression. With the assumption $\Omega >0$, we have 
\begin{align}
\text{Im} \Big[ \Pi_{jj}^{R}(\Omega)  \Big] = \pi \int_{\vec{k}} \text{Tr}&\Big( M_j P_{\vec{k},+1} M_j P_{\vec{k},-1}\Big)  \delta(\Omega - 2 E_{\bm{k}}), 
\end{align}
which can be evaluated by performing elementary integrals over $\bm{k}$. By combining with equation\eqref{def_op}, we finally obtain 
\begin{align}
\sigma_{jj}(\Omega)= \frac{\pi}{\Omega}\int_{\vec{k}} \text{Tr}&\Big( M_j P_{\vec{k},+1} M_j P_{\vec{k},-1}\Big)  \delta(\Omega - 2 E_{\bm{k}}). 
\label{fin_op}
\end{align}
Now we evaluate this for $M_x = ve \sigma^x$ and $M_y = A e k_y \sigma^y$ to calculate $\sigma_{xx}(\Omega)$ and $\sigma_{yy}(\Omega)$ for this anisotropic semimetal. 

Plugging $M_x = ve \sigma^x$ into equation\eqref{fin_op}, we have 
\begin{align}
\sigma_{xx}(\Omega) = \frac{e^2 v^2}{8\pi \Omega} \int dk_x dk_y \text{Tr}&\Big( \sigma_x P_{\vec{k},+1} \sigma_x P_{\vec{k},-1}\Big)  \times \delta(E_{\bm{k}} - \frac{\Omega}{2})
\end{align}
By the change of variables $\zeta_x = vk_x$ and $\zeta_y = A k_y^2$ and then performing the trace, we have 
\begin{align}
\sigma_{xx}(\Omega) = \frac{e^2 v}{16\pi \sqrt{A} \Omega} \int d\zeta_x d\zeta_y 
 &\delta(\sqrt{\zeta_x^2 + \zeta_y^4} - \frac{\Omega}{2}) \times \Big(1-\frac{\zeta_x^2 - \zeta_y^4}{\zeta_x^2 + \zeta_y^4}\Big)
\end{align}
Performing the change of the variables $\zeta_x = \Omega x$ and $\zeta_y = \sqrt{\Omega} y$, we find 
\begin{align}
\sigma_{xx}(\Omega) = \frac{e^2 v}{16\pi \sqrt{A}} \frac{I_{xx}}{\sqrt{\Omega}}
\end{align}
with the constant $I_{xx}$ defined as following
\begin{align}
I_{xx} &= \int dx dy ~\delta(\sqrt{x^2 + y^4} - \frac{1}{2}) \Big(1+ \frac{y^4-x^2}{y^4+x^2}\Big),\nonumber\\ 
&\approx 2.47 
\end{align}

We can plug $M_y  = A e k_y \sigma^y$ to calculate $\sigma_{yy}(\Omega)$. After the straightforward calculation, we find 
\begin{align}
\sigma_{yy}(\Omega) = \frac{e^2 \sqrt{A}}{4\pi v} I_{yy} \sqrt{\Omega}, 
\end{align}
where $I_{yy}$ is a constant 
\begin{align}
I_{yy} &= \int dx dy ~\delta(\sqrt{x^2 + y^4} - \frac{1}{2}) \Big(1+ \frac{x^2-y^4}{y^4+x^2}\Big), \nonumber\\ 
&\approx 4.94
\end{align}

In summary, we found anisotropic scalings in the optical conductivities along $\hat{x}$ and $\hat{y}$ directions 
\begin{align}
\sigma_{xx} (\Omega) \propto 1/\sqrt{\Omega}, ~ \sigma_{yy} (\Omega) \propto \sqrt{\Omega}, 
\end{align}

\section{Independence of Cutoff Scheme at QCP}
In this supplementary material, we show clearly that the RG flow equations near the QCP in \ref{RGNewFixedPoint}, i.e., Eq(8) in main text and Eq(69) of \ref{RGNewFixedPoint}, are cutoff scheme independent. To show this, we first note that the only non-trivial renormalization to the bare action \eqref{FinalRG1} is from the fermion self-energy, and thus we only need to show the independence of the cutoff scheme in the renormalization of the velocity $v$ and the mass $A$, which can be calculated from 
\begin{align}
-D_\Lambda \Sigma_f (\bm{k},\omega) = - \Lambda \frac{d}{d\Lambda} \Sigma_f (\bm{k},\omega). 
\end{align}
We introduce a soft cutoff function $J(\frac{q_y}{\Lambda}) = J(- \frac{q_y}{\Lambda})$ which is a smooth function (which is monotonically decreasing for $q_y >0$) satisfying 
\begin{align}
J(0) = 1, \quad J(\infty) = 0.  
\end{align} 
Then the fermion self-energy can be written as 
\begin{align}
-\Sigma_f(\bm{k}, \omega) = \frac{e^2}{2} \int_{\bm{q}} \frac{H(\bm{q}+\bm{k})}{E(\bm{q}+\bm{k})} g_b (\bm{q}) J(\frac{q_y + k_y}{\Lambda}) J(\frac{q_y}{\Lambda}). 
\end{align}
We will explicitly demonstrate that the renormalizations to the velocity $v$ and mass $A$ are independent of the detailed form of the function $J(\cdot)$, i.e., the flow equations near the QCP (see \ref{RGNewFixedPoint}),   
\begin{align}
&\frac{d}{dl} \log v= z-z_1 + \alpha f_1[\bar{\kappa}] \nonumber\\ 
&\frac{d}{dl} \log A= z-2 + \alpha f_2[\bar{\kappa}] \nonumber\\
\label{Mom2}
\end{align} 
with 
\begin{align}
&f_1[\bar{\kappa}] = \int^{\infty}_0 dx \frac{1}{1+\bar{\kappa} \sqrt{x}} \frac{1}{(x^2+1)^{3/2}}\nonumber\\
&f_2[\bar{\kappa}] = \int^{\infty}_0 dx \frac{1}{1+\bar{\kappa} \sqrt{x}} \frac{x^2(x^2-5)}{(x^2+1)^{5/2}}, 
\end{align}
are independent of the precise form of the cutoff function $J(\cdot)$. Note that the flow equations in \ref{RGNewFixedPoint} are originally obtained by using the hard cutoff function. 

\textbf{A. Correction to the velocity $v$}: we first calculate the correction to the velocity. We only need to calculate 
\begin{align}
-\Sigma_f (\bm{k}, \omega) = \frac{e^2}{2} \int_{\bm{q}} \frac{H(\bm{q}+\bm{k})}{E(\bm{q}+\bm{k})} g_b (\bm{q}) J^2 (\frac{q_y}{\Lambda}), 
\label{exper}
\end{align}
with $\bm{k} = (k_x, 0)$. Expanding in $\frac{H(\bm{q}+\bm{k})}{E(\bm{q}+\bm{k})}$ to $O(k_x)$, we find 
\begin{align}
-\Sigma_f (\bm{k}, \omega) = v k_x \sigma^x \times I_x, 
\end{align}
where 
\begin{align}
I_x = \frac{e^2}{2\pi^2}\int^{\infty}_0 dq_x  \int^{\infty}_0 dq_y &\Big( \frac{A^2 q_y^4}{[v^2 q_x^2 + A^2 q_y^4]^{3/2}}   \frac{1}{q_y + \kappa \sqrt{q_x}}  J^2 (\frac{q_y}{\Lambda}) \Big). 
\end{align}
We now perform the derivative $D_\Lambda$ on the self-energy to find 
\begin{align}
-D_\Lambda[\Sigma_f (\bm{k}, \omega)] = v k_x \sigma^x \times D_\Lambda[I_x].
\end{align} 
Hence we calculate
\begin{align}
D_\Lambda[I_x] = \frac{e^2}{2\pi^2}& \int^{\infty}_0 dq_x  \int^{\infty}_0 dq_y \Big( \frac{A^2 q_y^4}{[v^2 q_x^2 + A^2 q_y^4]^{3/2}}   \frac{1}{q_y + \kappa \sqrt{q_x}}  D_\Lambda[J^2 (\frac{q_y}{\Lambda})] \Big). 
\end{align}
To evaluate the integral, we first perform the change of the variable $y = \frac{q_y}{\Lambda}$ to find 
\begin{align}
D_\Lambda[I_x] =  - \frac{e^2}{2\pi^2} \Lambda \times& \int^{\infty}_0 dq_x  \int^{\infty}_0 dy \Big( \frac{A^2 \Lambda^4 y^4}{[v^2 q_x^2 + A^2 \Lambda^4 y^4]^{3/2}}  \frac{1}{\Lambda y + \kappa \sqrt{q_x}}   y \frac{d}{dy}[ J^{2}(y) ]\Big). 
\end{align}
Next we scale out $q_x \to q_x y$ to find 
\begin{align}
&D_\Lambda[I_x] = - \frac{e^2}{2\pi^2}  \Lambda  \int^{\infty}_0 dq_x \Big( \frac{A^2 \Lambda^4}{[v^2 q_x^2 + A^2 \Lambda^4]^{3/2}}  \frac{1}{\Lambda + \kappa \sqrt{q_x}}\Big)  \int^{\infty}_0 dy \frac{d}{dy} [J^2(y)], 
\end{align}
where the integral over $y$ can be performed analyticall to find 
\begin{align}
&D_\Lambda[I_x] =  \frac{e^2}{2\pi^2}  \Lambda  \int^{\infty}_0 dq_x \Big( \frac{A^2 \Lambda^4}{[v^2 q_x^2 + A^2 \Lambda^4]^{3/2}}  \frac{1}{\Lambda + \kappa \sqrt{q_x}}\Big), 
\end{align}
which is nothing but the momentum-shell integral for the renormalization of the velocity $v$ obtained in \eqref{Mom} of \ref{RGNewFixedPoint}, i.e., $I_1'$ in Eq.\eqref{Mom}. Notice that the expression is manifestly independent of the precise form of $J(\cdot)$. After successive change of variables as in \ref{RGNewFixedPoint}, we can show that the same integral $f_1[\bar{\kappa}]$ in Eq.\eqref{Mom2} appears in the renormalization of the velocity $v$. 

\textbf{B. Correction to the mass $A$}: we next calculate the correction to the mass $A$. After the straightforward series expansion of \eqref{exper} for $\bm{k} = (0, k_y)$ to $O(k_y^2)$ and then taking $D_\Lambda$, we find  
\begin{align}
-D_\Lambda[\Sigma_f(\bm{k},\omega)] =  Ak_y^2 \sigma^y (K_1 + K_2 + K_3), 
\end{align}   
where $\{ K_1, K_2, K_3 \}$ are the following integrals to be evaluated. 
\begin{align}
&K_1 = \frac{e^2}{2A}\int_{\bm{q}} g_b(\bm{q})\Big( \frac{1}{\Lambda^3} f(\bm{q})  \left(\Lambda J(\frac{q_y}{\Lambda})J''(\frac{q_y}{\Lambda}) + \frac{q_y}{2}  J'(\frac{q_y}{\Lambda})J''(\frac{q_y}{\Lambda})+ \frac{q_y}{2}  J(\frac{q_y}{\Lambda})J'''(\frac{q_y}{\Lambda})\right)\Big),  
\end{align}
in which $f(\bm{q}) = \frac{Aq_y^2}{\sqrt{v^2q_x^2 + A^2 q_y^4}}$ and we have used the notation $f'(x) = \frac{d}{dx} f(x)$ and similarly for $f''(x), f'''(x)$. For $K_2$, we have 
\begin{align}
&K_2 = \frac{e^2}{2A}\int_{\bm{q}} g_b(\bm{q})\Big( \frac{1}{\Lambda^3} f'(\bm{q}) \left(\Lambda^2 J(\frac{q_y}{\Lambda})J'(\frac{q_y}{\Lambda}) + q_y \Lambda  [J'(\frac{q_y}{\Lambda})]^2 + q_y \Lambda  J(\frac{q_y}{\Lambda})J''(\frac{q_y}{\Lambda})\right)\Big),  
\end{align}
and, for $K_3$, we have 
\begin{align}
&K_3 = \frac{e^2}{2A}\int_{\bm{q}}   \frac{g_b(\bm{q})}{\Lambda} f''(\bm{q})  q_y J(\frac{q_y}{\Lambda})J'(\frac{q_y}{\Lambda}). 
\end{align}
Performing the change of the variable $q_y = \Lambda y$ and then scaling $q_x \to q_x y$, we can show that 
\begin{align}
& K_1 \propto \int^{\infty}_0 dy \frac{d}{dy}[y^2 J(y)J''(y)] =  y^2 J(y)J''(y) \Big|^{\infty}_0 =0, \nonumber\\ 
& K_2 \propto \int^{\infty}_0 dy \frac{d}{dy}[y J(y)J'(y)] =  y^2 J(y)J'(y) \Big|^{\infty}_0 =0. 
\end{align}
On the other hand, we find 
\begin{align}
K_3 = - \frac{e^2}{2\pi^2} \Lambda \int^{\infty}_0 dq_x & \frac{1}{\kappa \sqrt{q_x} + \Lambda}\frac{v^2 q_x^2 [v^2 q_x^2 -5 A^2 \Lambda^4]}{[v^2 q_x^2 + A^2 \Lambda^4]^{5/2}}  \times \int^{\infty}_0 dy \frac{d}{dy} [J^2(y)]. 
\end{align}
By performing the integral over $y$ explicitly, we find 
\begin{align}
K_3 =  \frac{e^2}{2\pi^2} \Lambda \int^{\infty}_0 dq_x & \frac{1}{\kappa \sqrt{q_x} + \Lambda}\frac{v^2 q_x^2 [v^2 q_x^2 -5 A^2 \Lambda^4]}{[v^2 q_x^2 + A^2 \Lambda^4]^{5/2}}, 
\end{align} 
which is nothing but the momentum-shell integral for the renormalization of the mass $A$ obtained in \eqref{Mom} of \ref{RGNewFixedPoint}, i.e., $I_2'$ in Eq.\eqref{Mom}. Notice that the expression is manifestly independent of the precise form of $J(\cdot)$. After successive change of variables as in \ref{RGNewFixedPoint}, we can show that the same integral $f_2[\bar{\kappa}]$ in Eq.\eqref{Mom2} appears in the renormalization of the mass $A$.

\section{Ward Identity and non-renormalization of Electric Charge}
Here, we discuss the Ward identity and its consequences in two dimensional systems with the Coulomb interaction, and we also discuss difference between our work and Isobe {\it et. al.}\cite{Isobe2015}. The effective action after redefinition $e\phi \rightarrow \phi$ is 
\begin{eqnarray}
\mathcal{S} &=& \int_{\bm{x},\tau} \psi^{\dagger} \big( (\partial_{\tau} +i\phi)+ \mathcal{H}(-i \nabla) \big) \psi    \int_{q, \omega} \frac{|q| }{2 e^2}|\phi(q,\omega)|^2 + \mathcal{S}_{ins}, \nonumber
\end{eqnarray} 
where $\mathcal{S}_{ins}$ represents an additional inserted operator. For the discussion of the Ward identity, the inserted operator part is not important but it becomes crucial to investigate the stability of the non-interacting fixed point. Note that the action has gauge-invariance under 
\begin{eqnarray}
\phi \rightarrow \phi + \partial_{\tau} \alpha(\tau), \quad \psi \rightarrow \psi e^{-i \alpha(\tau)}. \nonumber
\end{eqnarray}
  
After incorporating quantum fluctuation, the effective action becomes
\begin{align}
\mathcal{S}\to &\int (1+\delta_{\omega})\psi^{\dagger} \partial_{\tau} \psi + (1+\delta_v) \int i \phi \,  \psi^{\dagger}\psi  +\int \psi^{\dagger} \tilde{H}(-i \nabla) \psi + \int \frac{|q|}{2 e^2} |\phi|^2 + \mathcal{S}_{ins}, \nonumber
\end{align}
Notice that non-analytic dependence of the $\frac{1}{e^2} |\bm{q}|$ term in $\bm{q}$ prohibits its renormalization under integrating out the high-energy modes. Here we did not explicitly write out the renormalized Hamiltonian $\tilde{H}(-i \nabla)$ as it is not important in the discussion of the Ward identity. 

It is clear that the gauge invariance gives the Ward identity, 
\begin{eqnarray}
\delta_{\omega} = \delta_v. \nonumber
\end{eqnarray} 
Moreover the non-renormalizable $e^2$ gives that the correction terms are absorbed by redefining $\psi$, $\psi = \sqrt{Z_{\psi}} \psi_r $ with $Z_{\psi}^{-1}=1+\delta_{\omega}$, and parameters in the Hamiltonian $\tilde{H}(-i \nabla)$ such as velocity and effective mass. 

Therefore, in our calculation, the electric charge is always non-renormalizable but $e Z_{\psi}$ receives correction even though  it appears in higher loop calculation with the bare Coulomb interaction.
 
This is one main difference between work by Isobe {\it et. al.}\cite{Isobe2015} and ours. In their work, they claim electric charge (g in their notation) is renormalized but $g Z$ ($e Z_{\psi}$ in our notation) is not. 

Another related difference between our work here and theirs\cite{Isobe2015} is their emphasis on the wavefunction renormalization $Z = (1+ \frac{\partial \Sigma_f (\omega, \bm{k})}{\partial (i \omega) })^{-1}$. However, in the strong-coupling limit $\alpha N \to \infty$, it is not difficult to see that $\frac{\partial \Sigma_f (\omega, \bm{k})}{\partial (i \omega) }$ contains the log-squared divergence (see below for the detail), which originates from the poor screening of the Coulomb interaction at finite frequency with zero momentum. In this limit, the only effect of the Coulomb interaction is to rotate the phase of the fermions as well explained in Son\cite{son} and thus should not be taken as physical. Such log-squared dependence is also present in Isobe {\it et.al.}\cite{Isobe2015} in the limit $\alpha N \to \infty$ and taken to calculate the fermion wavefunction renormalization $Z$. 
They interpret the wavefunction renormalization as single-particle residue. But it cannot be adiabatically connected to one of the strong coupling fixed point due to the additional divergence.  

One can see another difference from insertion operators $S_{ins}$. As shown above, the perturbative RG with the bare Coulomb potential without the inserted operator gives log-squared divergence in the inverse effective mass correction. This log-squared correction indicates necessity of additional operators. In our calculation, we insert the operator
\begin{eqnarray}
\mathcal{S}_{ins} = \frac{\tilde{\kappa}}{2} \int (\nabla \phi)^2, \nonumber
\end{eqnarray}
which is natural in the Wilsonian RG,  
while they\cite{Isobe2015} insert an approximated polarization function even for weak coupling analysis,
\begin{eqnarray}
\mathcal{S}_{ins}' = \int\frac{-1}{e^2} \Pi(q,\omega)|\phi(q,\omega)|^2. \nonumber
\end{eqnarray}
We believe that the insertion of an approximated polarization function is questionable in the weak coupling limit since it is not controlled by the $\frac{1}{N_f}$ factor, which disappeared in their weak coupling analysis. 

Note that the insertion of the approximated polarization which is contributed by the fermion modes from all energy scales is valid in strong coupling limit since it is well controlled by the $\frac{1}{N_f}$ factor. In the current problem, the strong coupling limit is indeed controlled, and all physical quantities receives $\frac{1}{N_f}$ corrections which are well captured in Isobe et. al. and ours with qualitative difference from different approximation. However, as shown in the main text, the strong coupling fixed point becomes immediately unstable due to the anomalous dimension of the velocity, thus the validity of the large-$N_f$ expansion is not guaranteed in the infrared limit.  

We also believe that unusual cutoff (scale) dependence of physical quantities in their analysis (for example, $\log \alpha$ in RG equations) indicates their RG scheme is not conventional and it cannot be perturbatively obtained from the non-interacting theory. 
On contrary, in our analysis, all physical quantities receive logarithmic cutoff (scale) dependence as ones of quantum criticality described by renormalizable field theories. 

Apart from the differences, we find the ground state and its excitation in the IR limit are marginally stable, so-called marginal Fermi liquid, which is also obtained by Isobe {\it et. al.} \cite{Isobe2015} in spite of different renormalization of the inverse mass.

\subsection{Log-squared Divergence in Fermion Self-Energy}
Here we present the log-squared divergence in the strong-coupling limit, $\alpha N_f \to \infty$, where $g_{b}^{-1}(\bm{q}, \Omega) \to -\Pi(\bm{q},\Omega)$ by keeping the dependence in $\Omega$. 

Such log-squared divergence has been observed in the graphene case as discussed by Son\cite{son}. The fermion self-energy in Son is given by 
\begin{align}
\Sigma(\bm{p}, p_0) = \Sigma_0 \gamma_0 p_0 + \Sigma_1 \bm{\gamma}\cdot \bm{p}, 
\end{align}
where 
\begin{align}
\Sigma_0 = \frac{8}{N_f} \int \frac{d^2 \bm{q} dq_0}{(2\pi)^3} \frac{q_0^2 - \bm{q}^2}{(q^2)^{3/2} |\bm{q}|^2}, 
\label{son1}
\end{align}
with $q = |(q_0, \bm{q})|$, the size of the three-component vector $(q_0, \bm{q})$, which is not to be confused with the size of the spatial two-component vector $|\bm{q}| = |(q_x, q_y)|$. Naively, this expression is single-log divergent, but it is in fact log-squared divergent. The log-squared divergence is apparent in the limit $\frac{|\bm{q}|}{q_0} \to 0$ because Eq.\eqref{son1} can be approximated 
\begin{align}
\Sigma_0 &\propto \int \frac{d^2 \bm{q} dq_0}{(2\pi)^3} \frac{q_0^2}{(q_0^2)^{3/2} |\bm{q}|^2} \propto \int \frac{dq_0}{q_0} \times \int \frac{ d^2 \bm{q}}{|\bm{q}|^2}, 
\end{align}
which has the two sources of the logarithmic divergence, one from the integral over $q_0$ and the other from the integral over $\bm{q}$. 

We expect the similar log-squared divergence to be present if we keep the dependence of the polarization $\Pi(\bm{q}, \Omega)$ in $\Omega$. To exhibit this explicitly, we reconsider the fermion self-energy and the boson polarization in the strong-coupling limit $\alpha \to \infty$. Here we show the log-sqaured divergence in $\Sigma_0$ where 
\begin{align}
\Sigma_0 = -\frac{1}{i\omega}\text{Tr} [\Sigma_f(\bm{k}, \omega)].  
\end{align}
Here we work in the unit $v=A = 1$ for convenience and the clarity of the discussion. We start from the expression 
\begin{align}
\Sigma_0 \propto \int_{\bm{q},\Omega} \frac{\Omega^2 -E_{\bm{q}}^2}{(\Omega^2 + E_{\bm{q}})^2} \frac{1}{\Pi(\bm{q},\Omega)}, 
\end{align}
where $E_{\bm{q}} = (q_x^2 + q_y^4)^{1/2}$. In the limit $\Omega \gg E_{\bm{q}}$, we have 
\begin{align}
\Sigma_0 &\propto \int_{\bm{q},\Omega} \frac{1}{\Omega^2} \frac{1}{\Pi(\bm{q},\Omega)} \propto \int_{\Omega} ~\frac{1}{\Omega} \times \int_{\bm{q}} \frac{1}{a_x \frac{q_x^2}{\sqrt{|\Omega|}} + a_y q_y^2 \sqrt{|\Omega|}}, 
\label{polar}
\end{align}
where $a_x \approx 0.66$ and $a_y \approx 0.75$ and we have used the following form of the polarization 
\begin{align}
\Pi(\bm{q},\Omega) \propto \frac{1}{\Omega} \Big[ a_x \frac{q_x^2}{\sqrt{|\Omega|}} + a_y q_y^2 \sqrt{|\Omega|} \Big],  
\label{polarr}
\end{align}
which is asymptotically correct as far as $\Omega \gg E_{\bm{q}}$. Now in Eq.\eqref{polar}, we can perform the change of the variables $x = \frac{\sqrt{a_x}}{|\Omega|^{1/4}} q_x$ and $y = \sqrt{a_y} |\Omega|^{1/4} q_y$ to find 
\begin{align}
\Sigma_0 \propto \int \frac{d\Omega}{\Omega} \times \int \frac{dxdy}{x^2+y^2}, 
\end{align}
which clearly exhibits the log-sqaured divergence. 

Now we show how we obtained the asymptotic expression of the boson self-energy Eq.\eqref{polarr}. We start with the expression for the boson self-energy Eq.\eqref{Pol}. By performing the following change of the variables $k_x = |\Omega| x$, $k_y = |\Omega|^{1/2} y$, $X=\frac{q_x}{|\Omega|}$ and $Y = \frac{q_y}{\sqrt{|\Omega|}}$ (remember that we took $v=A =1$ for this subsection) and expanding $\Pi$ to the lowest orders in $X$ and $Y$, we find 
\begin{align}
\Pi(\bm{q}, \Omega) &\approx \sqrt{|\Omega|} \Big[a_x X^2 + a_y Y^2 \Big]\nonumber\\ 
&= \frac{1}{\Omega}\Big[ a_x\frac{q_x^2}{\sqrt{|\Omega|}} + a_y \sqrt{|\Omega|} q_y^2 \Big]. 
\end{align}


\begin{thebibliography}{10}
\expandafter\ifx\csname url\endcsname\relax
  \def\url#1{\texttt{#1}}\fi
\expandafter\ifx\csname urlprefix\endcsname\relax\def\urlprefix{URL }\fi
\providecommand{\bibinfo}[2]{#2}
\providecommand{\eprint}[2][]{\url{#2}}

\bibitem{sachdev}
\bibinfo{author}{Sachdev, S.}
\newblock \emph{\bibinfo{title}{Quantum phase transitions}}
  (\bibinfo{publisher}{Wiley Online Library}, \bibinfo{year}{2007}).

\bibitem{millis}
\bibinfo{author}{Millis, A.}
\newblock \bibinfo{title}{Effect of a nonzero temperature on quantum critical
  points in itinerant fermion systems}.
\newblock \emph{\bibinfo{journal}{Phys. Rev.  B}}
  \textbf{\bibinfo{volume}{48}}, \bibinfo{pages}{7183} (\bibinfo{year}{1993}).

\bibitem{hertz}
\bibinfo{author}{Hertz, J.~A.}
\newblock \bibinfo{title}{Quantum critical phenomena}.
\newblock \emph{\bibinfo{journal}{Phys. Rev.  B}}
  \textbf{\bibinfo{volume}{14}}, \bibinfo{pages}{1165--1184}
  (\bibinfo{year}{1976}).

\bibitem{Hasan2011}
\bibinfo{author}{Hasan, M.~Z.} \& \bibinfo{author}{Moore, J.~E.}
\newblock \bibinfo{title}{{Three-Dimensional Topological Insulators}}.
\newblock \emph{\bibinfo{journal}{Annu. Rev. Condens. Matter Phys.}}
  \textbf{\bibinfo{volume}{2}}, \bibinfo{pages}{55--78} (\bibinfo{year}{2011}).

\bibitem{KaneRev}
\bibinfo{author}{{Hasan}, M.~Z.} \& \bibinfo{author}{{Kane}, C.~L.}
\newblock \bibinfo{title}{{Colloquium: Topological insulators}}.
\newblock \emph{\bibinfo{journal}{Rev. Mod. Phys.}}
  \textbf{\bibinfo{volume}{82}}, \bibinfo{pages}{3045--3067}
  (\bibinfo{year}{2010}).

\bibitem{QiRev}
\bibinfo{author}{{Qi}, X.-L.} \& \bibinfo{author}{{Zhang}, S.-C.}
\newblock \bibinfo{title}{{Topological insulators and superconductors}}.
\newblock \emph{\bibinfo{journal}{Rev. Mod. Phys.}}
  \textbf{\bibinfo{volume}{83}}, \bibinfo{pages}{1057--1110}
  (\bibinfo{year}{2011}).

\bibitem{SenthilReview}
\bibinfo{author}{Senthil, T.}
\newblock \bibinfo{title}{Symmetry-protected topological phases of quantum
  matter}.
\newblock \emph{\bibinfo{journal}{Annu. Rev. Condens. Matter Phys.}}
  \textbf{\bibinfo{volume}{6}}, \bibinfo{pages}{299--324}
  (\bibinfo{year}{2015}).

\bibitem{moon1}
\bibinfo{author}{Moon, E.-G.}, \bibinfo{author}{Xu, C.}, \bibinfo{author}{Kim,
  Y.~B.} \& \bibinfo{author}{Balents, L.}
\newblock \bibinfo{title}{Non-fermi-liquid and topological states with strong
  spin-orbit coupling}.
\newblock \emph{\bibinfo{journal}{Phys. Rev.  Lett.}}
  \textbf{\bibinfo{volume}{111}}, \bibinfo{pages}{206401}
  (\bibinfo{year}{2013}).

\bibitem{savary}
\bibinfo{author}{Savary, L.}, \bibinfo{author}{Moon, E.-G.} \&
  \bibinfo{author}{Balents, L.}
\newblock \bibinfo{title}{New type of quantum criticality in the pyrochlore
  iridates}.
\newblock \emph{\bibinfo{journal}{Phys. Rev.  X}}
  \textbf{\bibinfo{volume}{4}}, \bibinfo{pages}{041027} (\bibinfo{year}{2014}).

\bibitem{yang}
\bibinfo{author}{Yang, B.-J.}, \bibinfo{author}{Moon, E.-G.},
  \bibinfo{author}{Isobe, H.} \& \bibinfo{author}{Nagaosa, N.}
\newblock \bibinfo{title}{Quantum criticality of topological phase transitions
  in three-dimensional interacting electronic systems}.
\newblock \emph{\bibinfo{journal}{Nat. Phys.}}  (\bibinfo{year}{2014}).

\bibitem{hong}
\bibinfo{author}{Jian, S.-K.} \& \bibinfo{author}{Yao, H.}
\newblock \bibinfo{title}{Correlated double-weyl semimetals with coulomb
  interactions: Possible applications to HgCr${}_{2}$Se${}_{4}$ and SrSi${}_{2}$}.
\newblock \emph{\bibinfo{journal}{Phys. Rev. B}} \textbf{\bibinfo{volume}{92}},
  \bibinfo{pages}{045121} (\bibinfo{year}{2015}).

\bibitem{lai2015}
\bibinfo{author}{Lai, H.-H.}
\newblock \bibinfo{title}{Correlation effects in double-weyl semimetals}.
\newblock \emph{\bibinfo{journal}{Phys. Rev. B}} \textbf{\bibinfo{volume}{91}},
  \bibinfo{pages}{235131} (\bibinfo{year}{2015}).

\bibitem{amaricci2015}
\bibinfo{author}{Amaricci, A.}, \bibinfo{author}{Budich, J.},
  \bibinfo{author}{Capone, M.}, \bibinfo{author}{Trauzettel, B.} \&
  \bibinfo{author}{Sangiovanni, G.}
\newblock \bibinfo{title}{First-order character and observable signatures of
  topological quantum phase transitions}.
\newblock \emph{\bibinfo{journal}{Phys. Rev.  Lett.}}
  \textbf{\bibinfo{volume}{114}}, \bibinfo{pages}{185701}
  (\bibinfo{year}{2015}).

\bibitem{son}
\bibinfo{author}{Son, D.}
\newblock \bibinfo{title}{Quantum critical point in graphene approached in the
  limit of infinitely strong coulomb interaction}.
\newblock \emph{\bibinfo{journal}{Phys. Rev.  B}}
  \textbf{\bibinfo{volume}{75}}, \bibinfo{pages}{235423}
  (\bibinfo{year}{2007}).

\bibitem{sheehy}
\bibinfo{author}{Sheehy, D.~E.} \& \bibinfo{author}{Schmalian, J.}
\newblock \bibinfo{title}{Quantum critical scaling in graphene}.
\newblock \emph{\bibinfo{journal}{Phys. Rev.  Lett.}}
  \textbf{\bibinfo{volume}{99}}, \bibinfo{pages}{226803}
  (\bibinfo{year}{2007}).

\bibitem{pickett1}
\bibinfo{author}{Pardo, V.} \& \bibinfo{author}{Pickett, W.~E.}
\newblock \bibinfo{title}{Half-metallic semi-dirac-point generated by quantum
  confinement in TiO${}_2$/VO${}_2$ nanostructures}.
\newblock \emph{\bibinfo{journal}{Phys. Rev.  Lett.}}
  \textbf{\bibinfo{volume}{102}}, \bibinfo{pages}{166803}
  (\bibinfo{year}{2009}).

\bibitem{pickett2}
\bibinfo{author}{Pardo, V.} \& \bibinfo{author}{Pickett, W.~E.}
\newblock \bibinfo{title}{Metal-insulator transition through a semi-dirac point
  in oxide nanostructures: VO${}_2$ (001) layers confined within TiO${}_2$}.
\newblock \emph{\bibinfo{journal}{Phys. Rev.  B}}
  \textbf{\bibinfo{volume}{81}}, \bibinfo{pages}{035111}
  (\bibinfo{year}{2010}).

\bibitem{pickett3}
\bibinfo{author}{Banerjee, S.}, \bibinfo{author}{Singh, R.},
  \bibinfo{author}{Pardo, V.} \& \bibinfo{author}{Pickett, W.}
\newblock \bibinfo{title}{Tight-binding modeling and low-energy behavior of the
  semi-dirac point}.
\newblock \emph{\bibinfo{journal}{Phys. Rev.  Lett.}}
  \textbf{\bibinfo{volume}{103}}, \bibinfo{pages}{016402}
  (\bibinfo{year}{2009}).

\bibitem{a1}
\bibinfo{author}{Kobayashi, A.}, \bibinfo{author}{Suzumura, Y.},
  \bibinfo{author}{Pi{\'e}chon, F.} \& \bibinfo{author}{Montambaux, G.}
\newblock \bibinfo{title}{Emergence of dirac electron pair in the
  charge-ordered state of the organic conductor $\alpha$-(BEDT-TTF)$_{2}$I${}_3$}.
\newblock \emph{\bibinfo{journal}{Phys. Rev.  B}}
  \textbf{\bibinfo{volume}{84}}, \bibinfo{pages}{075450}
  (\bibinfo{year}{2011}).

\bibitem{a2}
\bibinfo{author}{Hasegawa, Y.}, \bibinfo{author}{Konno, R.},
  \bibinfo{author}{Nakano, H.} \& \bibinfo{author}{Kohmoto, M.}
\newblock \bibinfo{title}{Zero modes of tight-binding electrons on the
  honeycomb lattice}.
\newblock \emph{\bibinfo{journal}{Phys. Rev.  B}}
  \textbf{\bibinfo{volume}{74}}, \bibinfo{pages}{033413}
  (\bibinfo{year}{2006}).

\bibitem{a3}
\bibinfo{author}{Suzumura, Y.}, \bibinfo{author}{Morinari, T.} \&
  \bibinfo{author}{Pi{\'e}chon, F.}
\newblock \bibinfo{title}{Mechanism of dirac point in $\alpha$ type organic
  conductor under pressure}.
\newblock \emph{\bibinfo{journal}{J. Phys. Soc. Jap.}}
  \textbf{\bibinfo{volume}{82}}, \bibinfo{pages}{023708}
  (\bibinfo{year}{2013}).

\bibitem{o1}
\bibinfo{author}{Lee, K.~L.} \emph{et~al.}
\newblock \bibinfo{title}{Ultracold fermions in a graphene-type optical
  lattice}.
\newblock \emph{\bibinfo{journal}{Phys. Rev.  A}}
  \textbf{\bibinfo{volume}{80}}, \bibinfo{pages}{043411}
  (\bibinfo{year}{2009}).

\bibitem{o2}
\bibinfo{author}{Wunsch, B.}, \bibinfo{author}{Guinea, F.} \&
  \bibinfo{author}{Sols, F.}
\newblock \bibinfo{title}{Dirac-point engineering and topological phase
  transitions in honeycomb optical lattices}.
\newblock \emph{\bibinfo{journal}{New J. Phys.}}
  \textbf{\bibinfo{volume}{10}}, \bibinfo{pages}{103027}
  (\bibinfo{year}{2008}).

\bibitem{o3}
\bibinfo{author}{Tarruell, L.}, \bibinfo{author}{Greif, D.},
  \bibinfo{author}{Uehlinger, T.}, \bibinfo{author}{Jotzu, G.} \&
  \bibinfo{author}{Esslinger, T.}
\newblock \bibinfo{title}{Creating, moving and merging dirac points with a
  fermi gas in a tunable honeycomb lattice}.
\newblock \emph{\bibinfo{journal}{Nature}} \textbf{\bibinfo{volume}{483}},
  \bibinfo{pages}{302--305} (\bibinfo{year}{2012}).

\bibitem{lee1}
\bibinfo{author}{Lee, S.-S.}
\newblock \bibinfo{title}{Low-energy effective theory of fermi surface coupled
  with u (1) gauge field in 2+ 1 dimensions}.
\newblock \emph{\bibinfo{journal}{Phys. Rev.  B}}
  \textbf{\bibinfo{volume}{80}}, \bibinfo{pages}{165102}
  (\bibinfo{year}{2009}).

\bibitem{max1}
\bibinfo{author}{Metlitski, M.~A.} \& \bibinfo{author}{Sachdev, S.}
\newblock \bibinfo{title}{Quantum phase transitions of metals in two spatial
  dimensions. i. ising-nematic order}.
\newblock \emph{\bibinfo{journal}{Phys. Rev.  B}}
  \textbf{\bibinfo{volume}{82}}, \bibinfo{pages}{075127}
  (\bibinfo{year}{2010}).

\bibitem{max2}
\bibinfo{author}{Metlitski, M.~A.} \& \bibinfo{author}{Sachdev, S.}
\newblock \bibinfo{title}{Quantum phase transitions of metals in two spatial
  dimensions. ii. spin density wave order}.
\newblock \emph{\bibinfo{journal}{Phys. Rev.  B}}
  \textbf{\bibinfo{volume}{82}}, \bibinfo{pages}{075128}
  (\bibinfo{year}{2010}).

\bibitem{mross1}
\bibinfo{author}{Mross, D.~F.}, \bibinfo{author}{McGreevy, J.},
  \bibinfo{author}{Liu, H.} \& \bibinfo{author}{Senthil, T.}
\newblock \bibinfo{title}{Controlled expansion for certain non-fermi-liquid
  metals}.
\newblock \emph{\bibinfo{journal}{Phys. Rev.  B}}
  \textbf{\bibinfo{volume}{82}}, \bibinfo{pages}{045121}
  (\bibinfo{year}{2010}).

\bibitem{nayak1}
\bibinfo{author}{Nayak, C.} \& \bibinfo{author}{Wilczek, F.}
\newblock \bibinfo{title}{Non-fermi liquid fixed point in 2+ 1 dimensions}.
\newblock \emph{\bibinfo{journal}{Nucl. Phys. B}}
  \textbf{\bibinfo{volume}{417}}, \bibinfo{pages}{359--373}
  (\bibinfo{year}{1994}).

\bibitem{nayak2}
\bibinfo{author}{Nayak, C.} \& \bibinfo{author}{Wilczek, F.}
\newblock \bibinfo{title}{Renormalization group approach to low temperature
  properties of a non-fermi liquid metal}.
\newblock \emph{\bibinfo{journal}{Nucl. Phys. B}}
  \textbf{\bibinfo{volume}{430}}, \bibinfo{pages}{534--562}
  (\bibinfo{year}{1994}).

\bibitem{gonzalez}
\bibinfo{author}{Gonz{\'a}lez, J.}, \bibinfo{author}{Guinea, F.} \&
  \bibinfo{author}{Vozmediano, M.}
\newblock \bibinfo{title}{Marginal-fermi-liquid behavior from two-dimensional
  coulomb interaction}.
\newblock \emph{\bibinfo{journal}{Phys. Rev.  B}}
  \textbf{\bibinfo{volume}{59}}, \bibinfo{pages}{R2474} (\bibinfo{year}{1999}).

\bibitem{arpes}
\bibinfo{author}{Siegel, D.~A.} \emph{et~al.}
\newblock \bibinfo{title}{Many-body interactions in quasi-freestanding
  graphene}.
\newblock \emph{\bibinfo{journal}{Proc. Nat. Acad.
  Sci.}} \textbf{\bibinfo{volume}{108}}, \bibinfo{pages}{11365--11369}
  (\bibinfo{year}{2011}).

\bibitem{biswas}
\bibinfo{author}{Biswas, R.~R.}, \bibinfo{author}{Sachdev, S.} \&
  \bibinfo{author}{Son, D.~T.}
\newblock \bibinfo{title}{Coulomb impurity in graphene}.
\newblock \emph{\bibinfo{journal}{Phys. Rev.  B}}
  \textbf{\bibinfo{volume}{76}}, \bibinfo{pages}{205122}
  (\bibinfo{year}{2007}).

\bibitem{hosur}
\bibinfo{author}{Hosur, P.}, \bibinfo{author}{Parameswaran, S.} \&
  \bibinfo{author}{Vishwanath, A.}
\newblock \bibinfo{title}{Charge transport in weyl semimetals}.
\newblock \emph{\bibinfo{journal}{Phys. Rev.  Lett.}}
  \textbf{\bibinfo{volume}{108}}, \bibinfo{pages}{046602}
  (\bibinfo{year}{2012}).

\bibitem{carpentier}
\bibinfo{author}{Carpentier, D.}, \bibinfo{author}{Fedorenko, A.~A.} \&
  \bibinfo{author}{Orignac, E.}
\newblock \bibinfo{title}{Effect of disorder on 2d topological merging
  transition from a dirac semi-metal to a normal insulator}.
\newblock \emph{\bibinfo{journal}{Euro. Phys. Lett.}}
  \textbf{\bibinfo{volume}{102}}, \bibinfo{pages}{67010}
  (\bibinfo{year}{2013}).

\bibitem{Isobe2015}
\bibinfo{author}{Isobe, H.}, \bibinfo{author}{Yang, B.-J.},
  \bibinfo{author}{Chubukov, A.}, \bibinfo{author}{Schmalian, J.} \&
  \bibinfo{author}{Nagaosa, N.}
\newblock \bibinfo{title}{Emergent non-fermi liquid at the quantum critical
  point of a topological phase transition in two dimensions}.
\newblock \emph{\bibinfo{journal}{arXiv preprint arXiv:1508.03781}}
  (\bibinfo{year}{2015}).

\end{thebibliography}
\end{document}